\documentclass[12pt]{article}

\usepackage{graphicx}

\newcommand{\be}{\begin{equation}}
\newcommand{\ee}{\end{equation}}
\newcommand{\bea}{\begin{eqnarray}}
\newcommand{\eea}{\end{eqnarray}}
\newcommand{\ba}{\begin{array}}
\newcommand{\ea}{\end{array}}

\newcommand{\slashs}[1]{\not{\!#1}}
\newcommand{\norsl}{\normalsize\sl}
\newcommand{\norsc}{\normalsize\sc}

\begin{document}

\title{A technique for loop calculations in non-Abelian gauge theories\\
--with application to five gluon amplitude--} 

\author{
\norsc  Yoshiaki YASUI\\
\norsl  RIKEN BNL Research Center\\
\norsl  Brookhaven National Laboratory\\
\norsl  Upton, NY, 11973, USA\\
e-mail:yasui@bnl.gov
}

\date{}
\maketitle

\begin{abstract}
A powerful tool for calculations in non-Abelian gauge theories is
obtained by combining the background field gauge, the
helicity basis and the color decomposition methods.  It has reproduced 
the one-loop calculation of the five-gluon amplitudes in QCD, is
applicable to electroweak processes and extendable to two-loop
calculations.
\end{abstract}



\section{Introduction}

In search of new physics beyond the standard model that might be hiding
in jet events at high energy collider experiments, it is important to
understand the back ground arising from the known physics of the
standard model.  Here the conventional perturbation methods, which were so
useful in establishing the standard model itself, face a new challenge. 
Calculations for the multi-jet processes often induce vast numbers of
Feynman diagrams.  And to make the matter worse, the complicated
structure of the non-Abelian gauge-theory vertices amplify the number of
terms in the intermediate stages of calculations so as to make them
prohibitively difficult even at the tree level.  In this paper we present
a combination of the background field gauge, color decomposition and
spinor helicity basis which makes loop calculations of such multi-jet
processes feasible.

Parke et.al. developed methods to simplify multi-gluon amplitudes 
in the tree level \cite{PARKE1}.  
In their methods, gluon amplitudes are decomposed into color-ordered 
subamplitudes \cite{color} which are described by the spinor 
helicity basis \cite{SHB}. 
These techniques came from the analogy to the string theory.  
The subamplitudes constructed by above methods are gauge invariant 
and they possess the important relationships which are well known 
as the dual Ward identity. 
Supersymmetry is also useful in the pQCD calculations\cite{PARKE2}.  

 The string-motivated technique was pushed one step
further by Bern and Kosower and their coworkers \cite{BDK1,BDK2,BDK3}.  
They introduced a new technique, which is called  the string-inspired  
method or Bern-Kosower Rule,  to compute the one-loop pQCD amplitudes.   
This technique is based on the technology of string theory.  The idea 
is that string amplitudes include the pQCD amplitudes in the infinite 
string tension limit.  Using this technique, they performed the one 
loop calculation for the process $gg \rightarrow gg$ as a non trivial 
example\cite{BDK1,BDK2}.  Their results agree with the results of the 
conventional  calculations obtained by Ellis and  Sexton\cite{ELLIS}.   
They also gave the first calculation of the one-loop amplitude for 
five external gluons\cite{BDK3}. 
It is one of the most difficult part in the calculation of the next 
to leading order contribution for the three jets production 
processes\cite{KILG}.  
No one has done it in the conventional Feynman diagram formula so far.

It is well known that, at the tree level, the string inspired method 
is connected with the Gervais-Neveu non-linear gauge\cite{DRM}. 
However, the relation between the one-loop level string motivated 
calculation and the conventional pQCD calculation is still subtle. 
Many people suggested that there is a relation 
between the string inspired method and the background field gauge\cite{BD}.
The background field method and/or gauge is another powerful method. 
In the early works by DeWitt, the background field method was formulated to compute the quantum corrections for the effective 
action without losing explicit gauge invariance\cite{DEWITT,HART}.
In this gauge, we can construct the gauge invariant  
effective action $\Gamma[B]$, which is invariant 
under the gauge transformation of the 
classical background field $B_\mu$.
It was developed by Abbot in the QCD case\cite{ABB}. 
He applied this method to compute the 1PI Green's functions. 
He also explained that the correct S-matrix is given  
from trees of 1PI Green's functions constructed in the 
background field gauge\cite{ABB2}. 
This means that the background field gauge allows us to 
calculate the S-matrix in a gauge invariant way. 

The background field gauge is a conventional 
method based on the Feynman diagram formula.  This method
possesses several advantages to carry out loop calculations 
of the QCD amplitudes. In the following we
demonstrate that this background field gauge combined with color
decomposition and spinor helicity techniques provide a powerful tool to
calculate loop amplitudes in pQCD.  To prove its effectiveness we
present an example of the one-loop calculation of the five-gluon
amplitudes.

The paper is organized as follows. 
A simple review of the background field method is given in section 2.
In section 3,
we also review the color decomposition technique and give the color order 
Feynman rules in the background field gauge. 
The helicity basis method is a popular technique in multi-jet analysis. 
We give the review of this method in the section 4.
In section 5 we discuss how to combined method is 
applies to the calculation of the five-gluon
amplitudes at one-loop level.
The section 6 is the conclusion.

\section{Background field gauge}

The idea of the background field method is to construct the gauge 
invariant effective action. 
In the background field method, the gauge field $A$ in the classical 
Lagrangian is split into the  classical background field $B$ and the 
quantum field $Q$.
\[
{\cal L}(A)={\cal L}(Q+B).
\]
For the pure gauge theory, 
${\cal L}(A)=-{1\over4}F^a_{\mu\nu}F^{a\mu\nu}$ and 
$F^a_{\mu\nu}$ is the filed strength of $A$.  
From the analogy of the conventional generating functional $Z$,
\[
{Z}[J] = \int {\cal D}A~\det{M}
\exp
i\left[ 
\int d^4 x \{{\cal L}(A) 
-{1\over 2\alpha}{G}\cdot {G} +J\cdot A\}\right],
\]
we can construct the background field 
generating functional $\tilde{Z}$ through the orthodox procedure 
of the path integral formula with the Lagrangian 
${\cal L}(Q+B)$, 
\[
\tilde{Z}[J,B] = \int {\cal D}Q~\det\tilde{M}
\exp
i\left[ 
\int d^4 x \{{\cal L}(Q+B) 
-{1\over 2\alpha}\tilde{G}\cdot \tilde{G} +J\cdot Q\}\right].
\]
Here, ${G}~(\tilde{G})$ term is the gauge fixing term and 
$\det{M}~(\det\tilde{M})$ is the Faddeev-Popov determinant. 
 Faddeev-Popov determinants are given by the derivative of 
gauge fixing term under the infinitesimal gauge transformation, 
\[
\delta{A}^a_\mu
=-f^{abc}\omega^b A^c_\mu+{1\over g}\partial_\mu \omega^a,
\]
for $M=\delta{G}/\delta\omega$, and, 
\[
\delta{Q}^a_\mu
=-f^{abc}\omega^b (Q+B)^c_\mu+{1\over g}\partial_\mu \omega^a,
\]
for $\tilde{M}=\delta\tilde{G}/\delta\omega$. 
We choose the gauge fixing condition for the background field 
generating functional as,
\be
\tilde{G}=D^B_\mu\cdot Q^{a\mu}
\equiv
\partial_\mu Q^{a\mu}+gf^{abc}B^b_\mu Q^{c\mu},
\label{eq:BG}
\ee
which is called the background field gauge. 
We also get the background field effective action 
by the Legendre transformation, 
\[
\tilde{\Gamma}[\tilde{Q},B]
=\tilde{W}[J,B]-\int d^4x J\cdot \tilde{Q},
\]
where, 
\[
\tilde{W}[J,B]=-i \ln \tilde{Z}[J,B]
~~~~~~~~~~~~~~~~~~
\tilde{Q}={\delta W\over \delta J}.
\]
The relation between the background field effective action 
$\tilde\Gamma$ 
and the conventional effective action $\Gamma$ is given by,
\[
\tilde{\Gamma}[\tilde{Q},B]=\Gamma[\tilde{Q}+B]_B.
\]
The index $B$ in the RHS refers to the $B$ dependence 
in the gauge fixing term,
\be
G_\mu^a(Q,B)=D^B_\mu\cdot(Q-B)^{a\mu}. 
\label{eq:BG2}
\ee
For the special case $\tilde{Q}=0$, we have the important 
relation\cite{ABB},
\[
\tilde{\Gamma}[0,B]=\Gamma[B]_B.
\]
The RHS is the conventional effective action which is calculated 
with the gauge fixing condition eq.(\ref{eq:BG2}).
It is invariant under the gauge transformation of $B$. 
$\tilde{\Gamma}[0,B]$ has no dependence on the $\tilde{Q}$, 
thus, 1PI diagrams reduced from this effective action have 
only background fields $B$'s as the  external legs. 
In other words, the gauge invariant effective action is calculated 
by summing up the vacuum diagrams in the presence of the 
classical background field $B$. 
Though the
background field gauge method provides a different set of Green's
functions than conventional methods with conventional gauges, the
correct S-matrix is constructed from trees of 1PI diagrams
\cite{ABB2}. 

In addition, 
the Feynman rules of the background fields gauge have 
a very simple structure. For example, the three point vertex 
of the one external and two internal gluons 
is given as,

\vspace{0.3cm}
\begin{minipage}[t]{.3\linewidth}
\resizebox{4cm}{4cm}{\includegraphics{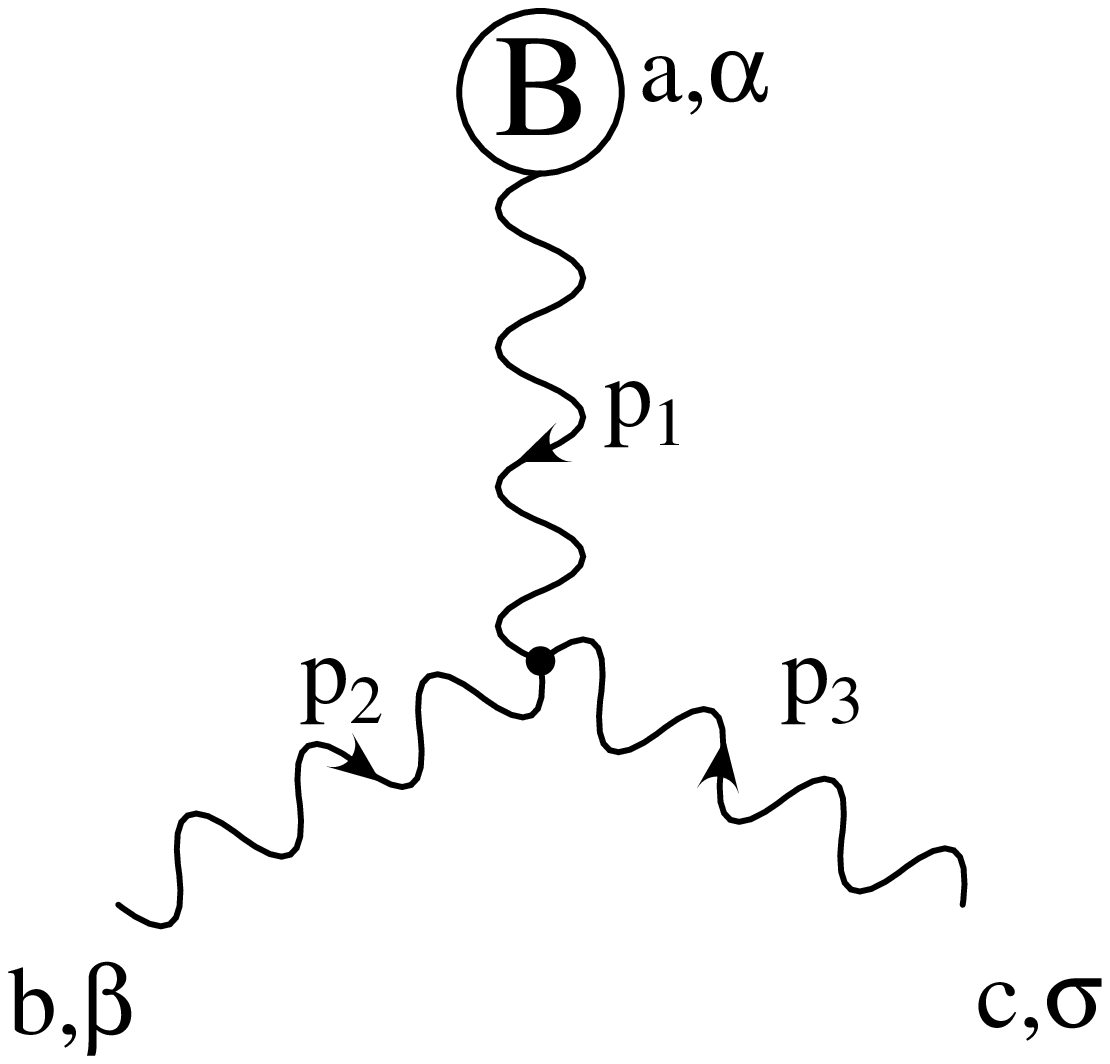} }
\end{minipage}
\hfill
\begin{minipage}{.7\linewidth}
\vspace{-4cm} 
\bea
&~&
g f^{abc}
\left[(p_2-p_3)_\alpha g_{\beta\sigma}
+(p_3-p_1+{1\over \xi}p_2)_{\beta}
g_{\alpha\sigma}\right.
\nonumber\\
&~&~~~~~~~~~~~~
\left.
+(p_1-p_2-{1\over \xi}p_3)_{\sigma}
g_{\alpha\beta}\right].
\nonumber
\eea
\end{minipage}\\
Here,  $\xi$ is the gauge fixing parameter for the 
internal quantum fields. 
If we choose the gauge fixing parameter as $\xi=1$, 
the above rule becomes, 
\[
igf^{abc}[(p_2-p_3)_\alpha g_{\beta\sigma}
+2p_{1\sigma}g_{\alpha\beta}
-2p_{1\beta}g_{\alpha\sigma}].
\]
We notice that only first $(p_2-p_3)$ term includes 
the internal momenta. 
Thus, only this term induces the integration momenta 
into the numerators of the Feynman integrals. 
In general, loop integrals which include integration momenta 
in their numerators induce 
a huge number of terms and make the calculation complicated. 
Thus, the background field method 
suppresses the number of the terms which appear in the intermediate 
stages of the loop calculations remarkably.
We will discuss this for the case of the five gluon vertex 
in the section 5.
Note, the same vertex structure also appears in the  
string inspired method. Bern and Dunbar discussed the relation between 
the string inspired method and the conventional field theory in 
ref.\cite{BD}.  
They pointed out that there is a mapping between the string motivated 
rules for the loop calculation and the Feynman rules of 
the background field gauge.

\section{Color decomposition}

The color decomposition is the technique which 
constructs color ordered gauge invariant subamplitudes 
in the SU(N) gauge theory. 
At the tree level, n-point 
gluon scattering amplitudes ${\cal M}_n$ for the SU(N) gauge theory 
can be decomposed into the subamplitudes $m$'s which are 
characterized single traces of the group matrices\cite{PARKE1}. 
It is well known as the Chan-Paton factor. 
\[
{\cal M}_n = \sum_{a_i \in S_n/Z_n}
 {\rm Tr}(T^{a_1} T^{a_2} \cdots T^{a_n})
m_n(p_{a_1}^{i_{a_1}},p_{a_2}^{i_{a_2}},\cdots ,p_{a_n}^{i_{a_n}}),
\]
where $p_j$ are external momenta, $i_j$ are helicity.
$T^a~(a=1,2,\cdots,N^2-1)$ are the matrices of the gauge 
group in the fundamental representation. $S_n/Z_n$ denotes 
the set of noncyclic permutations over $1,\cdots,n$.
Each subamplitudes have the independent color structures. 
Thus these color decomposed subamplitudes are gauge invariant.
In addition, 
$m_n(p_{a_1}^{i_{a_1}},p_{a_2}^{i_{a_2}},\cdots ,p_{a_n}^{i_{a_n}})$ 
is invariant under cyclic permutations of $p_j^{i_j}$. 
The subamplitudes also satisfy some important properties 
which is known as the Dual Ward identity\cite{PARKE1}. 

For the pure SU(N) gauge theory, 
we can consider U(N) theory instead of the SU(N) theory.
Since the U(1) mode decouples from the SU(N) mode, 
the U(1) contributions must automatically vanish in the final 
results of the SU(N) gluon amplitudes. 
However, U(N) group has the larger symmetry than the SU(N) group.
Thus presence of the U(1) mode simplify the intermediate stage 
of the calculations. 
$T^a$ ($a=1,2,\cdots,N^2-1$) are generators of SU(N) group  
for the fundamental representation. 
In this paper, we use the normalization condition of the generators as  
${\rm Tr}(T^a T^b)=\delta^{ab}/2$. 
(This normalization is different from the Bern et.al.'s one 
in the factor $\sqrt{2}$. ) It satisfies the 
relation between the adjoint representation 
and the fundamental representation,
\[
{\rm Tr}[T^a~,[T^b,T^c]] = {i\over2} f^{abc}.
\]
We introduce the U(1) mode  $T^0$,
\[
T^0_{ij}={\delta_{ij}\over \sqrt{2N}}.
\]
The factor ${1\over \sqrt{2N}}$ is convention which 
does not change the normalization condition of $T$'s.
The algebra is modified as following from SU(N) to 
SU(N)$\times$ U(1) theory.
\[
\sum_{a=1}^{N^2-1} T^a_{ij} T^a_{kl} 
={1\over2}\left(\delta_{il}\delta_{jk}
-{1\over N}\delta_{ij}\delta_{kl}\right)
~~\Longrightarrow~~
\sum_{a=0}^{N^2-1} T^a_{ij} T^a_{kl} 
={1\over2}\left(\delta_{il}\delta_{jk}\right).
\]
Here we sum up over the indices $c =0, 1,2, \cdots ,N^2-1$
for the SU(N)$\times$ U(1) .
The Casimir factor is given by,
\[
\sum_{a=1}^{N^2-1}T^a T^a ={N^2-1\over 2 N}
~~\Longrightarrow~~
\sum_{a=0}^{N^2-1}T^a T^a = {N\over2}
\]
We also obtain the following simple formulas 
of the Fierz identities, 
\bea
\sum_{b=0}^{N^2-1}
Tr(T^{a_1}T^{a_2} \cdots T^{a_m} T^b)(T^b~ T^{a_{m+1}} \cdots T^{a_n})
&=& {1\over 2} Tr(T^{a_1} T^{a_2} 
\cdots T^{a_m} T^{a_{m+1}} \cdots T^{a_n}),
\nonumber\\
\sum_{b=0}^{N^2-1}
Tr(T^{a_1} \cdots T^{a_m} T^b~ T^{a_{m+1}}\cdots T^{a_n} ~T^b)
&=&
{1\over2}Tr(T^{a_1} \cdots T^{a_m})Tr(T^{a_{m+1}}\cdots T^{a_n}).
\nonumber\\
\label{eq:form2}
\eea

For the one loop level gluon amplitudes, 
double-trace components also appear. 
For example, the color decomposition of the one-loop 
five-gluon amplitude is given by\cite{BK},
\bea
{\cal M}_n &=& 
\sum_{a_i \in S_5/Z_5}
 {\rm Tr}(T^{a_1} T^{a_2} \cdots T^{a_5})
m_{5,1}(p_{a_1}^{i_{a_1}},\cdots ,p_{a_5}^{i_{a_5}})
\nonumber\\
&+&
\sum_{a_i \in S_5/S_{5;2}}
 {\rm Tr}(T^{a_1}T^{a_2}){\rm Tr}(T^{a_3}T^{a_4}T^{a_5})
m_{5,3}(p_{a_1}^{i_{a_1}},\cdots ,p_{a_5}^{i_{a_5}})
\nonumber
\eea
Double-trace components $m_{5,3}$ are related with 
$m_{5,1}$ via the decoupling equation. 
This relation can be easily derived from the string theory\cite{BK}. 
We also can derive it in the straightforward way 
with using the U(N) Fierz identities eq.(\ref{eq:form2}). 
So, we only need to consider the single-trace part $m_{5,1}$. 
Single-trace parts $m_{5,1}$ are a leading contribution of 
the large $N$ expansion in the U(N) and SU(N) gauge theory. 
Leading order contributions of the large $N$ expansion 
are given directly with using the color ordered Feynman rules. 
In addition, only a color ordered subset of all the Feynman diagrams 
is required. 
In other word, we only have to consider the topologically 
independent diagrams and their cyclic permutations 
on the external color charges to calculate the amplitudes $m_{5,1}$. 
Thus we reach the color-ordered Feynman rules
of the background field gauge as summarized in the following.
The color factors $(a,b,\cdots)$ 
denote ${\rm Tr}(T^a,T^b,\cdots)$. 
We also give the diagrams of the five gluon vertices 
at one-loop level in Appendix.

\vspace{1cm}
\begin{minipage}[t]{.3\linewidth}
\input epsf.sty
\epsfxsize=4cm
\epsffile{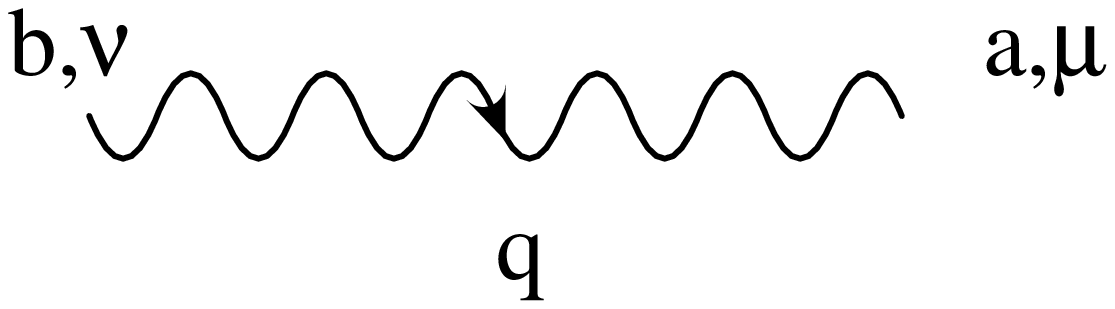}
\end{minipage}
\hfill
\begin{minipage}{.7\linewidth}
\vspace{-2cm}
\[
-{i\delta_{ab}g_{\mu\nu}\over
  q^2+i\varepsilon}
\]
\end{minipage}

\vspace{0.5cm} 
\begin{minipage}[t]{.3\linewidth}
\input epsf.sty
\epsfxsize=4cm
\epsffile{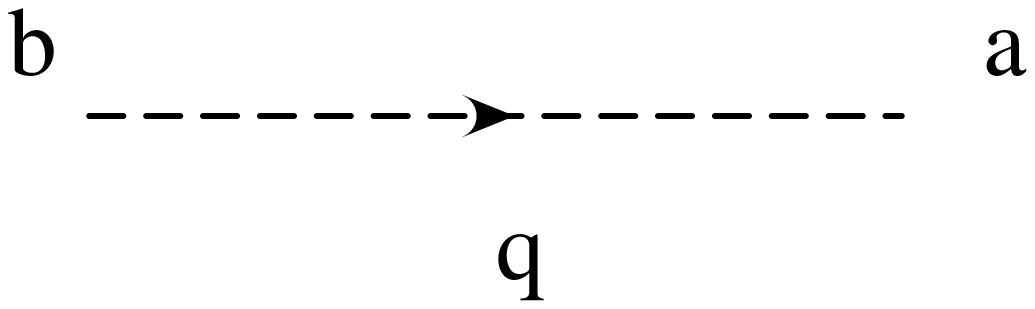}
\end{minipage}
\hfill
\begin{minipage}{.7\linewidth}
\vspace{-2cm}
\[
{i\delta_{ab}\over
  q^2+i\varepsilon}
\]
\end{minipage}

\begin{minipage}[t]{.3\linewidth}
\input epsf.sty
\epsfxsize=3.5cm
\epsffile{bgw3.ps} 
\end{minipage}
\hfill
\begin{minipage}{.7\linewidth}
\vspace{-4cm} 
\[
-2ig~(abc)
[(p_2-p_3)_\alpha g_{\beta\sigma}
+2p_{1\sigma}g_{\alpha\beta}
-2p_{1\beta}g_{\alpha\sigma}]
\]
\end{minipage}

\begin{minipage}[t]{.3\linewidth}
\input epsf.sty
\epsfxsize=3.5cm
\epsffile{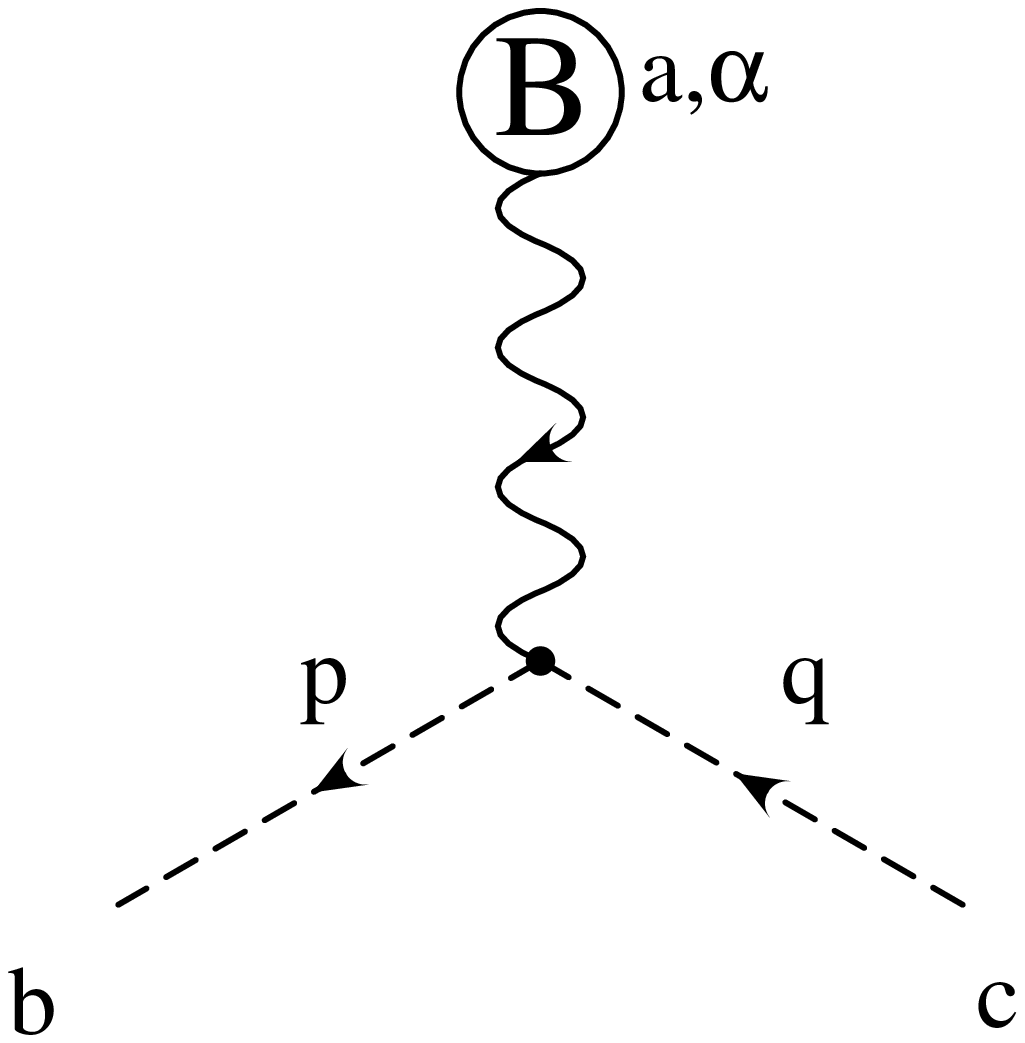} 
\end{minipage}
\hfill
\begin{minipage}{.7\linewidth}
\vspace{-4cm}
\[
-2ig~(abc)(p+q)_\alpha
\]
\end{minipage}

\begin{minipage}[t]{.3\linewidth}
\input epsf.sty
\epsfxsize=4cm
\epsffile{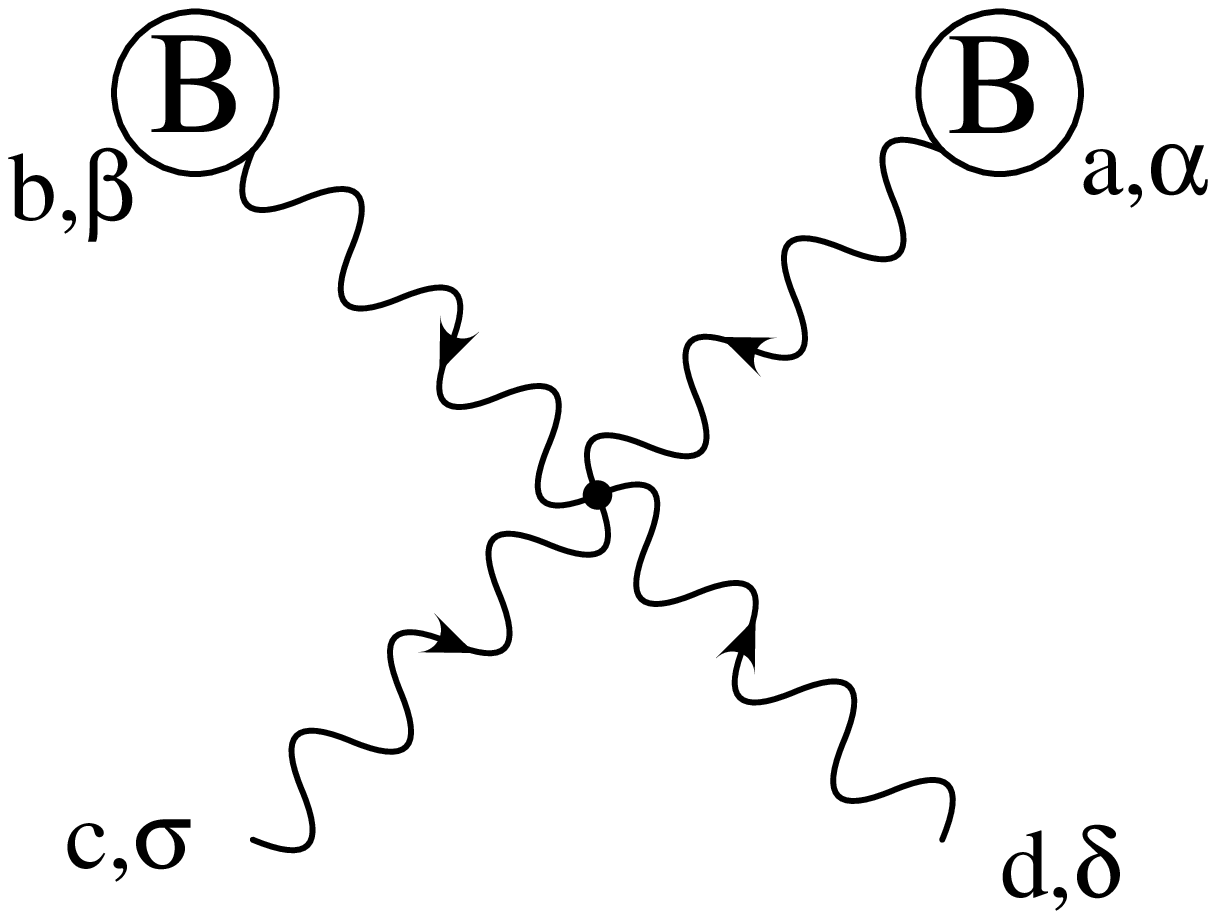} 
\end{minipage}
\hfill
\begin{minipage}{.7\linewidth}
\vspace{-3cm}
\[
-4ig^2
(abcd)(g_{\alpha\delta}g_{\beta\sigma}
-g_{\alpha\sigma}g_{\beta\delta}
+{1\over2}g_{\alpha\beta}g_{\sigma\delta})
\]
\end{minipage}

\vspace{0.3cm}
\begin{minipage}[t]{.3\linewidth}
\input epsf.sty
\epsfxsize=4cm
\epsffile{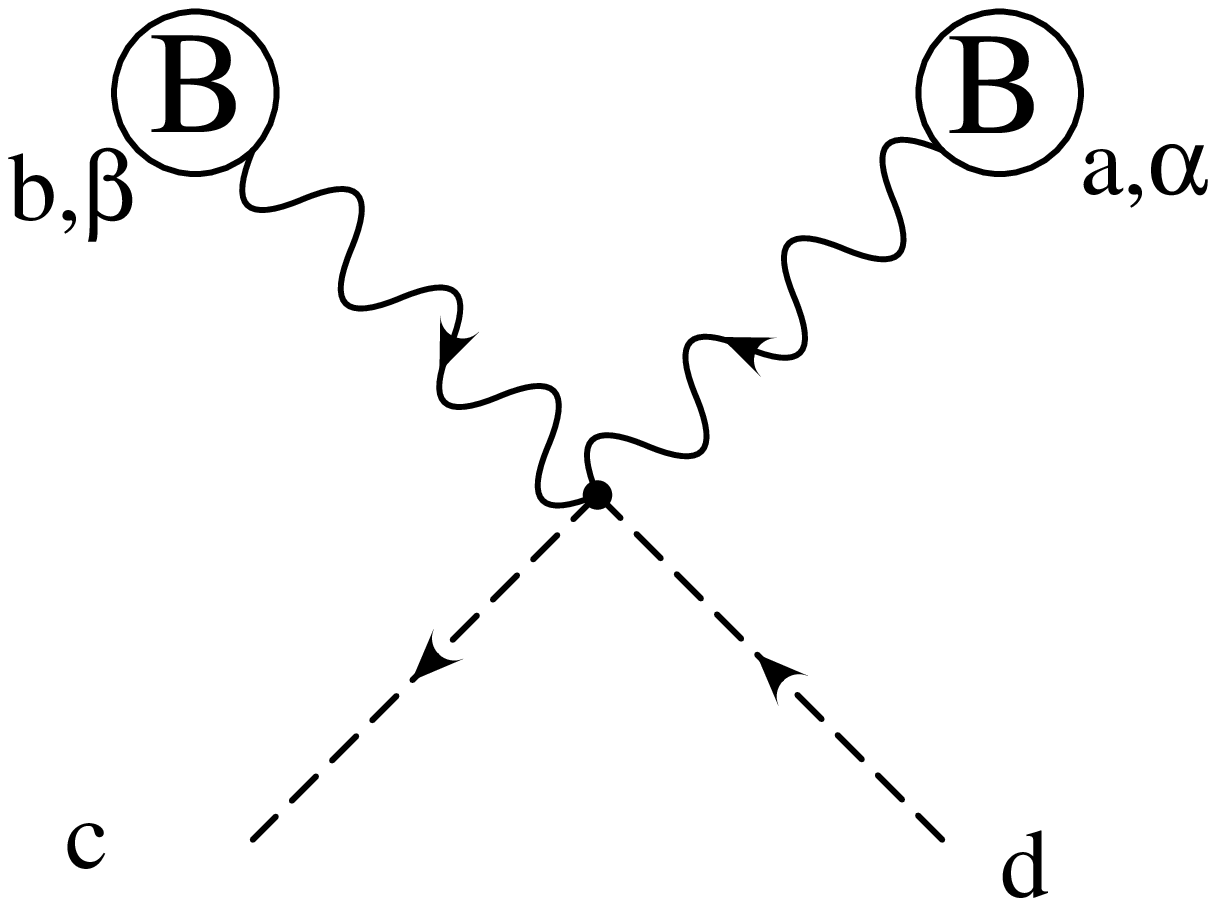} 
\end{minipage}
\hfill
\begin{minipage}{.7\linewidth}
\vspace{-3cm}
\[
2ig^2~(abcd)g_{\alpha\beta}
\]
\end{minipage}

\section{Spinor Helicity basis}

It is well known that 
the helicity basis method is often useful in the tree 
level calculations\cite{PARKE1,SHB}. 
It is useful in the loop calculation, too\cite{KOS}. 
In this method, we calculate matrix elements 
in which polarizations of the external fields are 
characterized by the spinor helicity basis. 
Here we introduce the well known notations 
on the helicity basis of a spinor field $\psi$ as,
\[
\langle q^{\pm}|
={1\over2}\overline{\psi}(q)(1\mp \gamma_5),
~~~~~~~~~~~~~~~
|\bar{q}^{\pm}\rangle 
={1\over2}(1\pm \gamma_5)\psi(\bar{q}).
\]
Here all fermions are massless.  
The normalization condition is given by,
\[
\langle p|\gamma_\mu|p\rangle =2p_\mu.
\]
It is convenient to introduce the following 
notation of spinor products,
\[
\langle pq\rangle =\langle p^-|q^+\rangle 
,~~~~~
[pq]=\langle p^+|q^-\rangle 
,~~~~~
[pq]\langle qp\rangle
=s_{pq}=2p\cdot q
.
\]

The most remarkable advantage of this formula is that we can also 
describe external gauge fields by using the spinor helicity basis. 
Thus, the complicated tensor structure of the multi-gluon amplitudes 
are replaced into the calculation of the Dirac algebra. 
The polarization vectors may be written in terms of  massless spinors 
$|p^\pm\rangle$ and $|k^\pm\rangle$,
\be
\varepsilon^{\pm}(p,k)
=
\pm{\langle p^\pm|\gamma_\mu|k^\pm\rangle 
\over \sqrt{2}\langle k^\mp|p^\pm\rangle }
\label{eq:pol}
\ee
where $p$ is the gauge boson momentum, 
$k$ is the arbitrary momentum which satisfies 
$k^2=0$. 
We call this momentum $k$ as the {\it reference} momentum. 

The final results for physical observable do not depend on the 
{\it reference} momentum because a change in the reference momentum 
is equivalent to a gauge transformation:
\[
\varepsilon^{+}(p,k)_\mu
\rightarrow
\varepsilon^{+}(p,k')_\mu-\sqrt{2}
{\langle kk'\rangle \over\langle kp\rangle \langle k'p\rangle }p_\mu.
\]
This means we have freedom in choosing an appropriate reference 
momentum for any gauge invariant subset of the full amplitude, 
such as a gauge invariant color-ordered subamplitude.

The polarization vectors defined by eq.(\ref{eq:pol})  satisfy 
not only the equation of motion, 
\[
p_\mu \varepsilon^{\pm}(p,k)^\mu=0,
\]
but also, 
\[
k_\mu \varepsilon^{\pm}(p,k)^\mu=0.
\]
Using the Fierz identity,
\be
\langle p^+|\gamma_\mu|q^+\rangle 
\langle k^-|\gamma^\mu|l^-\rangle 
=2[pl]\langle kq\rangle , 
\label{eq:hel1}
\ee
and a symmetric property,
\[
\langle p^+|\gamma_\mu|q^+\rangle 
=
\langle q^-|\gamma_\mu|p^-\rangle, 
\]
it is easy to 
show the following relations, 
\bea
\varepsilon^{\pm}(p,k)\cdot \varepsilon^{\pm}(p,k')&=&0
\nonumber\\
\varepsilon^{\pm}(p,k)\cdot \varepsilon^{\pm}(p',k)&=&0. 
\nonumber
\eea
In the actual calculations, we take advantage of the fact that 
choices of the reference momenta are not unique. 
Some of the good choices of the reference momenta  make it easy 
to use the above identities which reduce the number of terms in 
the calculations. The case of the four gluon amplitudes are shown 
in the ref.\cite{BDK2}. 

For the five gluon case, in this paper, we choose the reference momenta 
of the helicity $+$ gluons in the $m_5(+,+,+,+,+)$ amplitude as,
 \[
\epsilon_{\mu}(l_i,k_i=l_{i+1})
={
\langle l_i|\gamma_\mu|l_{i+1}\rangle
\over
\sqrt{2}\langle l_{i+1}l_i\rangle},
\]
where $l_i$ ($k_i$) is a $i$-th external gluon (reference) momentum. 
Using this expression, we can replace the complicated tensor structures 
into the calculation of the Dirac algebra. 
For example, contraction of the momentum $l$'s and helicity $+$ 
external gluon fields are, 
\bea
l_i^\alpha l_j^\beta l_k^\sigma l_l^\delta l_m^\rho
\times
\varepsilon_\alpha^{1+}
\varepsilon_\beta^{2+}
\varepsilon_\sigma^{3+}
\varepsilon_\delta^{4+}
\varepsilon_\rho^{5+}
&=&
{
\langle 1^+|{i}|2^+\rangle
\langle 2^+|{j}|3^+\rangle
\langle 3^+|{k}|4^+\rangle
\langle 4^+|{l}|5^+\rangle
\langle 5^+|{m}|1^+\rangle
\over(\sqrt{2})^5
\langle 21 \rangle
\langle 32 \rangle
\langle 43 \rangle
\langle 54 \rangle
\langle 15 \rangle
}
\nonumber\\
&~&
\nonumber\\
&=&
{Tr(
1i2j3k4l5m
P_+
)
\over(\sqrt{2})^5
\langle 21 \rangle
\langle 32 \rangle
\langle 43 \rangle
\langle 54 \rangle
\langle 15 \rangle
},
\nonumber
\eea
where $\varepsilon_\mu^{i+}\equiv\varepsilon_\mu^{+}(l_i,k=l_{i+1})$, 
$\langle i|j|k\rangle\equiv\langle l_i|\slashs{l}_j|l_k\rangle$ 
and $\langle ij \rangle\equiv \langle l_i l_j\rangle$, 
$P_+ \equiv{1\over2}(1+\gamma_5)$, 
$Tr(ij\cdots)\equiv Tr(\slashs{l}_i\slashs{l}_j\cdots)$. 
Here we used the identities, 
\be
\langle p|k|q\rangle = [pk]\langle kq\rangle
\label{eq:<p|k|q>}
\ee
and
\bea
{1\over2}
Tr(\slashs{p}_1\slashs{p}_2\cdots\slashs{p}_{2n}
(1+\gamma_5))
&=&
[p_1p_2]\langle p_2 p_3 \rangle \cdots
\langle p_{2n} p_1\rangle
\nonumber\\
{1\over2}
Tr(\slashs{p}_1\slashs{p}_2\cdots\slashs{p}_{2n}
(1-\gamma_5))
&=&
\langle p_1p_2\rangle [ p_2 p_3 ]\cdots
[p_{2n} p_1].
\label{eq:TRACE}
\eea
We also choose $(k_1,k_2,k_3,k_4,k_5)=
(l_2,l_3,l_4,l_5,l_2)$ for the $m_5(-,+,+,+,+)$ case.
Of curse, the choice of the reference momentum is not unique. 
Other choice is possible and may be more efficient efficient. 

In this paper, we would like to apply the helicity basis method to 
the one-loop calculation. 
To carry out the Feynman integral, 
the dimensional regularization is efficient. 
In the conventional dimensional regularization scheme, 
all gluon polarizations are dealt with in the 
$4-2\varepsilon$ dimension. 
On the other hand, in the helicity basis method, 
gluon polarization vectors are defined in 4 dimensions, 
because, the spinor helicity basis is only well defined 
in 4 dimensions. 
Thus, we need some modification on the regularization scheme. 
The Four Dimensional Helicity(FDH) scheme is one of the solutions 
which is effective in the helicity basis method\cite{BDK2}. 
In this scheme, all gluon polarizations (of observed and unobserved) 
are dealt with in 4 dimensions. Thus all gluons have 2 helicity states. 
The 'tHooft-Veltman scheme is also applicable. But, in this scheme, 
unobserved gluon (virtual, soft and collinear) polarizations are kept 
in  $4-2\varepsilon$ dimensions and we only treat the 
observed gluon polarizations in 4 dimensions. 

Before discussing the one-loop calculation in the next section, 
we give some comments on the tree-level results. 
For the tree level calculation, supersymmetry is useful\cite{PARKE2}.
Supersymmetric Ward identities 
show that maximal helicity violating(MHV) 
and next helicity violating(NHV) 
multi-gluon amplitudes of the pure supersymmetric QCD vanish,
\[
m^{SUSY}_n(l_1^\pm,l_2^+,\cdots,l_n^+)=0,
\]
where $l_i$ are external momentums and indices $\pm$ denote helicities.
In the supersymmetric theory, additional super particles, eg.   
scalar gluons and gluinos, contribute to the amplitudes. 
However, if we assume the R-symmetry, tree level amplitudes include 
no exotic couplings between gluons and additional particles. 
Thus, the non-supersymmetric MHV and NHV amplitudes 
for general $n$ gluon vertices also vanish in the tree level,  
\[
m^{tree}_n(l_1^\pm,l_2^+,\cdots,l_n^+)=0.
\]
This simple results ensure that MHV and NHV amplitudes 
in the one loop level must be infrared finite. 

For the other combinations of the helicities, amplitudes do not vanish. 
But using the spinor helicity formula, 
we obtain very simple expressions of the color ordered helicity 
amplitudes,   
\[
m^{tree}_n(p_1^+,p_2^+,\cdots,p_i^-,\cdots,p_j^-,\cdots,p_n^+)=
ig^{n-2}(\sqrt{2})^n{\langle ij \rangle ^4
\over
\langle 12 \rangle\langle 23 \rangle
\cdots 
\langle n1 \rangle}
,
\]
where $\langle ij \rangle=\langle p_i p_j \rangle$.
One-loop amplitudes for these helicities induce the Infrared
singularity\cite{BDK3}.

\section{Five gluon vertex example} 

In this  section, we  would like to  demonstrate the  background field
gauge  combined   with  the  helicity  basis  method   and  the  color
decomposition  is a powerful  tool for  the one-loop  calculation.  We
perform the  one-loop calculation of  the five gluon amplitudes  as an
example.   The  one-loop  level   five  gluon  amplitudes  were  first
calculated by  Bern, Dixon  and Kosower with  using the  new technique
which is  called the  string inspired method.   Here we show  that the
background  field   gauge  is  powerful,  too.    Combination  of  the
background field gauge, the color decomposition and the helicity basis
method  simplifies the  calculation  enough and  make  it feasible  to
compute the one-loop five gluon amplitudes in the straightforward way.

To carry out  this calculation, we need the  information of the tensor
type  Feynman  integrals for  the  pentagon  diagram.   We follow  the
technology  on the  Feynman  integral calculation  which discussed  in
references  \cite{BDK4}.  Here  we consider  the general  form  of the
dimensionally regulated massless Pentagon integral,
\be
{\cal I}_5[k^{\mu}k^{\nu}k^{\rho}\cdots
]
\equiv
\int{d^Dk\over(2\pi)^D}
{\mu^{2\epsilon}k^{\mu} k^{\nu}k^{\rho}\cdots\over
k^2(k-p_1)^k(k-p_2)^2(k-p_3)^2(k-p_4)^2},
\label{eq:GENERAL}
\ee
where
$D=4-2\varepsilon$, 
$\mu$ is the dimensional regulation scale parameter, 
$p_i=\sum_{j=1}^il_j$ and $l_i$ are external momenta. 
We also introduce the following notation to simplify 
the calculation of the tensor integrals, 
\be
I_5[k^{\mu}k^{\nu}k^{\rho}\cdots]
\equiv
i (4\pi)^{2-\epsilon}
\mu^{-2\epsilon}
{\cal I}_5[k^{\mu}k^{\nu}k^{\rho}\cdots].
\label{eq:kint}
\ee

For example, using the color ordered Feynman rules presented in section 3, 
explicit form of the color ordered 
gluonic pentagon integral, is given by,
 
\vspace{0.3cm}
\begin{minipage}[t]{.4\linewidth}
\input epsf.sty
\epsfxsize=6cm
\epsffile{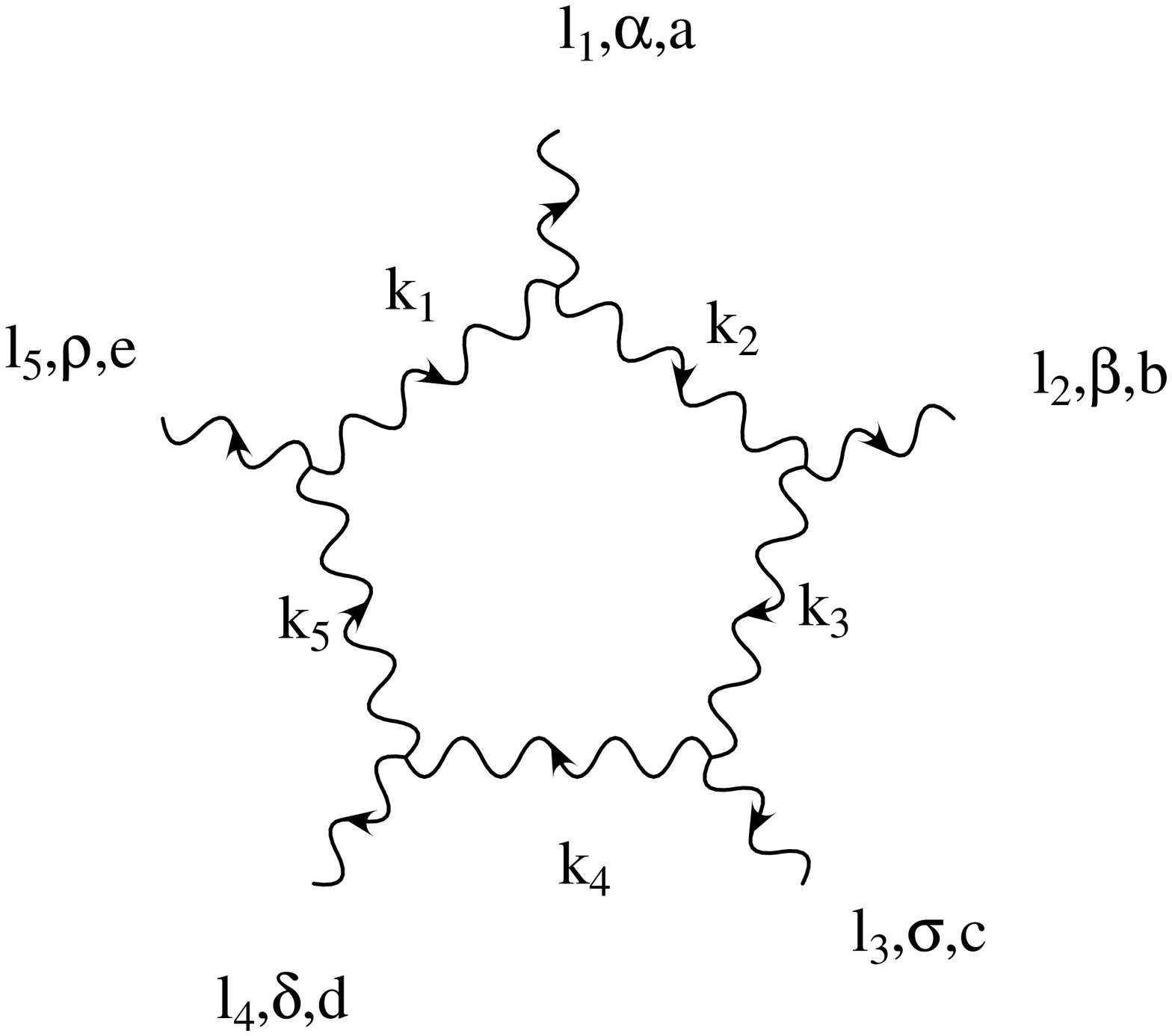} 
\end{minipage}
\hfill
\begin{minipage}{.6\linewidth}
\vspace{-4cm} 
{\small{
\bea
I^g_5
&=&i^5 N~{Tr(abcde)}
\nonumber\\
&\times&
\int {d^4k\over (2\pi)^4}
{(-i2g)^5\over k_1^2 k_2^2 k_3^2 k_4^2 k_5^2}
\nonumber\\
&\times&[{k_{1\alpha}} g_{\mu\nu}
        +l_{1\mu}g_{\alpha\nu}
        -l_{1\nu}g_{\alpha\mu}]
\nonumber\\
&\times&[{k_{2\beta}} g_{\nu\lambda}
        +l_{2\nu}g_{\beta\lambda}
        -l_{2\lambda}g_{\beta\nu}]
\nonumber\\
&\times&[{k_{3\sigma}} g_{\tau\lambda}
        +l_{3\lambda}g_{\sigma\tau}
        -l_{3\tau}g_{\sigma\lambda}]
\nonumber\\
&\times&[{k_{4\delta}} g_{\xi\tau}
        +l_{4\tau}g_{\delta\xi}
        -l_{4\xi}g_{\delta\tau}]
\nonumber\\
&\times&[{k_{5\rho}} g_{\mu\xi}
        +l_{5\xi}g_{\rho\mu}
        -l_{5\mu}g_{\rho\xi}].
\nonumber
\eea
}}
\end{minipage}
where $Tr(ab\cdots)=Tr(T^aT^b\cdots)$, 
$k_i$ are internal momenta 
and $l_i$ are external gluon momentums.
We notice that only the first term of the each vertices 
include the integration momenta $k$. 
Thus, the amplitude $I_5^g$ is decomposed into 
\bea
I_5^g
&=&
F_0 
I_5[k_1^\alpha k_2^\beta k_3^\sigma k_4^\delta k_5^\rho]
+
F_1^\rho
I_5[k_1^\alpha k_2^\beta k_3^\sigma k_4^\delta]
+
F_2^{(1)\delta\rho}
I_5[k_1^\alpha k_2^\beta k_3^\sigma]
+
F_2^{(2)\sigma\rho}
I_5[k_1^\alpha k_2^\beta k_4^\delta]
\nonumber\\
&+&
F_3^{(1)\sigma\delta\rho}
I_5[k_1^\alpha k_2^\beta ]
+
F_3^{(2)\beta\delta\rho}
I_5[k_1^\alpha k_3^\sigma ]
+
F_4^{\beta\sigma\delta\rho}
I_5[k_1^\alpha]
+
F_5^{\alpha\beta\sigma\delta\rho}
I_5[1]
\nonumber\\
&+&
\mbox{cyclic permutation } 
\nonumber
\eea
If we use the conventional gauge, 
like the covariant gauge, 
other terms also include the integration momenta $k$.
Thus, more combinations on ${I}_5[k'\mbox{s}]$ appear.

Integrating over the momentum $k$  
after the Feynman parameterization, 
the momentum integral (\ref{eq:kint}) is rewritten by 
the Feynman parameter integral. The tensor structure is 
decomposed by the terms of momentums $l's$ and 
metric tensors $g_{\mu\nu}$. 
For example, 
the tensor integrals which appear  
in the pentagon integral $I^g_5$, is described by,
\bea 
I_5[k^{\mu_1}]&=&\sum_{i=1}^4 p_i^{\mu_1}I_5[a_{i+1}]
\nonumber\\
I_5[k^{\mu_1}k^{\mu_2}]
&=&\sum_{i,j=1}^4 p_i^{\mu_1}p_j^{\mu_2}I_5[a_{i+1}a_{j+1}]
-{1\over2}g^{\mu_1 \mu_2}I_5^{D=6-2\epsilon}
\nonumber\\
I_5[k^{\mu_1}k^{\mu_2}k^{\mu_3}]
&=&\sum_{i,j,k=1}^4 p_i^{\mu_1}p_j^{\mu_2}p_k^{\mu_3}
I_5[a_{i+1}a_{j+1}a_{k+1}]
-{1\over2}\sum_i \{gp_i\}^{\mu_1 \mu_2 \mu_3}
I_5^{D=6-2\epsilon}[a_{i+1}],
\nonumber\\
I_5[k^{\mu_1}k^{\mu_2}k^{\mu_3}k^{\mu_4}]
&=&\sum_{i,j,k,l=1}^4 p_i^{\mu_1}p_j^{\mu_2}p_k^{\mu_3}p_l^{\mu_4}
I_5[a_{i+1}a_{j+1}a_{k+1}a_{l+1}]
\nonumber\\
&~&
-{1\over2}\sum_{i,j}\{gp_i p_j \}^{\mu_1 \mu_2 \mu_3 \mu_4}
I_5^{D=6-2\epsilon}[a_{i+1} a_{j+1}]
+{1\over4}\{gg\}^{\mu_1 \mu_2 \mu_3 \mu_4}
I_5^{D=8-2\epsilon},
\nonumber\\
I_5[k^{\mu_1}k^{\mu_2}k^{\mu_3}k^{\mu_4}k^{\mu_5}]
&=&\sum_{i,j,k,l,m=1}^4 p_i^{\mu_1}p_j^{\mu_2}p_k^{\mu_3}
p_l^{\mu_4}p_m^{\mu_5}
I_5[a_{i+1}a_{j+1}a_{k+1}a_{l+1} a_{m+1}]
\nonumber\\
&~&
-{1\over2}\sum_{i,j,k}\{gp_i p_j p_k\}^{\mu_1 \mu_2 \mu_3 \mu_4 \mu_5}
I_5^{D=6-2\epsilon}[a_{i+1} a_{j+1} a_{k+1}]
\nonumber\\
&~&
+{1\over4}\sum_i \{ggp_i\}^{\mu_1 \mu_2 \mu_3 \mu_4 \mu_5}
I_5^{D=8-2\epsilon}[a_{i+1}],
\nonumber\\
&~&
\label{eq:tensor}
\eea
where $p_i=\sum_{j=1}^il_j$, 
$I_5[a_i,\cdots]$ are Feynman parameter integrals 
which are same as the notation introduced in ref.\cite{BDK4} 
and $a_i$ are Feynman parameters, 
\[
I_n[a_{i_1}\cdots a_{i_m}]
\equiv
\Gamma(n-2+\epsilon)
\int_0^1d^na_i\delta(1-\sum_i a_a)
{a_{i_1}\cdots a_{i_m}\over 
{\cal D}(a_i)^{n-2+\epsilon}}
\]
\bea
{\cal D}(a_i)&\equiv& 
  \sum_{i.j=1}^n D_{ij}a_i a_j.
~~~~~~~~~~
D_{ij}\equiv{1\over2}(-P_{ij}^2)
\nonumber\\
P_{ij}&\equiv&p_{j-1}-p_{i-1}
=l_i+l_{i+1}+\cdots+l_{j-1}~~~~
{\rm for} i<j.
\label{eq:genfey}
\eea
The $\{gp_i\cdots\}^{\mu_1\cdots}$ denotes the summation over 
the all possible permutations of Lorentz indices, for example, 
\[
\{gp_i\}^{\mu_1 \mu_2 \mu_3}
=g^{\mu_1 \mu_2}p_i^{\mu_3}
+g^{\mu_3 \mu_1}p_i^{\mu_2}
+g^{\mu_2 \mu_3}p_i^{\mu_1}.
\]
To perform the Feynman parameter integrals, 
we use the dimensional regulated formula discussed 
by Bern et.al.\cite{BDK4}.
The idea of this formula is to construct algebraic equations for 
n-point one-loop integrals. 
For the $D=4-2\epsilon$ scalar n-point integral case,  
we have,
\be
I_n[a_i]
={1\over2}
\left\{
\sum_{j=1}^n D^{-1}_{ij} {I}^{(i)}_{n-1}[1]
+(n-5+2\epsilon)
c_i{I}^{D=6-2\epsilon}_{n}[1]
\right\},
\label{eq:In[a]}
\ee
where, 
\[
c_i = \sum_{j=1}^n D_{ij}^{-1}, 
\]
and $D_{ij}$ is defined in eq.(\ref{eq:genfey}). 
$I_{n-1}^{(k)}$ is the n-1 point integral corresponds to 
removing the propagater parameterized by $a_k$ from the 
integral $I_{n}$. 
Using the identity $\sum_i^n I_n[a_i]=I_n[1]$, we have, 
\be
{I}_n[1]
={1\over 2}\left\{
\sum_{i=1}^n c_i {I}^{(i)}_{n-1}[1]
+(n-5+2\epsilon)
c_0{I}^{D=6-2\epsilon}_{n}[1]
\right\},
\label{eq:In[1]}
\ee
where, 
$c_0 = \sum_{i=1}^n c_{i}.$
Since $I^{D=6-2\epsilon}_5$ is finite, 
$\epsilon\rightarrow 0$ limit for $n=5$ case reproduces 
the  Melrose and van Neerven et.al.'s result\cite{VAN},
\[
{I}_5[1]
={1\over 2}
\sum_{i=1}^5 c_i {I}^{(i)}_{4}[1]
+{\cal O}(\epsilon)
.
\]
We notice that the scalar pentagon integral in the 4 dimension 
is obtained by a sum of the five box integrals. 
For the calculation of the tensor type Feynman integral, 
we also need the information of the Feynman integrals 
$I_n[a_i,a_j\cdots]$. 
By the changing of the integration variables 
in eq.(\ref{eq:genfey}) from $a_i$ to $u_i$\cite{thooft}, 
\[
a_i
={\alpha_i u_i  \over \sum_{j=1}^n \alpha_j u_j},
~~~~~~~~
a_n={\alpha_n (1-\sum_{j=1}^{n-1}u_j)
\over \sum_{j=1}^n \alpha_j u_j},
\]
it is very easy to show, 
\be
I_n[a_{i_1}a_{i_2}\cdots a_{i_m}]
={\Gamma(n-3-m+2\epsilon)\over \Gamma(n-3+2\epsilon)}
{\cal A}_n
  \alpha_{i_1}\alpha_{i_2}\cdots \alpha_{i_m}
{\partial\over\partial \alpha_{i_1}}
{\partial\over\partial \alpha_{i_2}}
\cdots
{\partial\over\partial \alpha_{i_m}}
\left(
{
I_n[1]\over {\cal A}_n}
\right),
\label{eq:In[aaa]}
\ee
where ${\cal A}_n=\prod_{j=1}^n\alpha_j$. 
From the eq.(\ref{eq:In[a]}) 
and eq.(\ref{eq:In[1]}),the scalar pentagon integral $I_5[1]$ and 
the one parameter pentagon integral $I_5[a_i]$ are described by 
$D=4-2\epsilon$ box integrals in the ${\cal O}(\epsilon^0)$.
Explicit form of the scalar box integral is, 
\[
{I}^{(k)}_4
=2\gamma_\Gamma {\cal A}_4^{(k)}\left[
{(\alpha_{k+2}\alpha_{k-2})^\epsilon\over \epsilon^2}
+{\rm Li}_2\left(1-{\alpha_{k+1}\over \alpha_{k+2}}\right)
+{\rm Li}_2\left(1-{\alpha_{k-1}\over \alpha_{k-2}}\right)
-{\pi^2\over 6}\right]
+{\cal O}(\epsilon),
\]
where ${\cal A}_4^{(k)}=\prod_{j(\neq k)=1}^4\alpha_j$. 
${\rm Li}_2(Z)$ is the Spence function, 
\[
{\rm Li}_2(Z)=-\int_0^Z {dx\over x} \log(1-x).
\]
The parameters $\alpha_i$ satisfy $s_{ii+1}=1/\alpha_i\alpha_{i+2}$. 
Thus, pentagon integrals $I_5[1]$ and $I_5[a_i]$ are given as,
\bea
{I}_5[1] 
&=&
\gamma_\Gamma\sum_{j=1}^5\alpha_j^{1+2\epsilon}
A_5 \left[
{1\over \epsilon^2}
+2{\rm Li}_2\left(1-{\alpha_{j+1}\over\alpha_j}\right)
+2{\rm Li}_2\left(1-{\alpha_{j-1}\over\alpha_j}\right)
-{\pi^2\over 6}\right]
+{\cal O}(\epsilon)
\nonumber\\
{I}_5[a_i] 
&=&
\gamma_\Gamma
A_5 \alpha_i\left[
{\alpha_i^{2\epsilon}\over \epsilon^2}
+2{\rm Li}_2\left(1-{\alpha_{j+1}\over\alpha_j}\right)
+2{\rm Li}_2\left(1-{\alpha_{j-1}\over\alpha_j}\right)
-{\pi^2\over 6}\right]
+{\cal O}(\epsilon)
\nonumber\\
\eea
The pentagon integrals which have two Feynman parameters 
in the numerator is calculated by the following relation, 
\be
{I}_5^{D=4-2\epsilon}[a_i a_j]
={\alpha_i \gamma_i\over \hat{\Delta}_n}
{I}_5^{D=4-2\epsilon}[a_j]
+
\sum_k 
{{\cal R}_{ik}\over 2}
{I}^{\stackrel{(k)}{D=4-2\epsilon}}_{4}
[a_j]
\nonumber\\
+
{{\cal R}_{ij}\over 2}{I}_5^{D=6-2\epsilon}[1].
\label{eq:Rn,n-1(1)}
\ee
where, 
\bea
\hat{\Delta}_5
&=&
\sum_{i=1}^5(\alpha_i^2-2\alpha_i \alpha_{i+1}
+2\alpha_i\alpha_{i+2})
=\gamma_5\gamma_2+\gamma_1\gamma_3+\gamma_2\gamma_4+
\gamma_3\gamma_5+\gamma_4\gamma_1
\nonumber\\
\gamma_i&=&{1\over2}{\partial\hat{\Delta}_5\over \partial \alpha_i}
=
\alpha_{i-2}-\alpha_{i-1}+\alpha_{i}
-\alpha_{i+1}+\alpha_{i+2},
\nonumber\\
{\cal R}_{ij}&=&
\alpha_i \alpha_j
(\eta_{ij}-{\gamma_i \gamma_j \over \hat{\Delta}_5})
~~~~~~~~~~~
\eta_{ij}=
\left\{
\begin{array}{ll}
-1 &(i=j\pm1)\\
+1 &({\rm others})
\end{array}
\right.
\nonumber
\eea
The box integral $I_4^{(k)}[a_i]$ in eq.(\ref{eq:Rn,n-1(1)}) is, 
\bea
{I}_4^{\stackrel{(k)}{D=4-2\epsilon}}[a_i]
&=&\gamma_\Gamma {\cal A}_4^{(k)}\alpha_i\left[
\delta_{k,i-1}\left\{-{1\over\epsilon^2}{
\alpha_{i-2}^\epsilon (\alpha_i^\epsilon-\alpha_{i+1}^\epsilon)\over
\alpha_i-\alpha_{i+1}}
+{\alpha_{i+2}L_k \over \hat{\Delta}_5-\gamma_k^2}\right\}\right.
\nonumber\\
&~&+
\delta_{k,i-2}\left\{{1\over\epsilon^2}\
\left({\alpha_{i-1}^\epsilon\alpha_{i+1}^\epsilon\over\alpha_i}
+{
\alpha_{i+2}^\epsilon (\alpha_{i-1}^\epsilon-\alpha_{i}^\epsilon)\over
\alpha_{i-1}-\alpha_{i}}\right)
+{(\alpha_{i+2}-\alpha_{i+1})L_k \over \hat{\Delta}_5-\gamma_k^2}\right\}
\nonumber\\
&~&+
\delta_{k,i+2}\left\{{1\over\epsilon^2}\
\left({\alpha_{i-1}^\epsilon\alpha_{i+1}^\epsilon\over\alpha_i}
+{
\alpha_{i-2}^\epsilon (\alpha_{i+1}^\epsilon-\alpha_{i}^\epsilon)\over
\alpha_{i+1}-\alpha_{i}}\right)
+{(\alpha_{i-2}-\alpha_{i-1})L_k \over \hat{\Delta}_5-\gamma_k^2}\right\}
\nonumber\\
&~&+\left.
\delta_{k,i+1}\left\{-{1\over\epsilon^2}{
\alpha_{i+2}^\epsilon (\alpha_i^\epsilon-\alpha_{i-1}^\epsilon)\over
\alpha_i-\alpha_{i-1}}
+{\alpha_{i-2}L_k \over \hat{\Delta}_5-\gamma_k^2}\right\}
\right],
\nonumber
\eea
where 
\[
L_i=Li_2(1-{\alpha_{i+1}\over \alpha_{i+2}})
+Li_2(1-{\alpha_{i-1}\over \alpha_{i-2}})
+{\rm ln}{\alpha_{i+1}\over \alpha_{i+2}}
{\rm ln}{\alpha_{i-1}\over \alpha_{i-2}}
-{1\over 6}\pi^2.
\]
$D=6-2\epsilon$ integrals are given from 
the analytic continuation $D=4-2\epsilon$ to $D=6-2\epsilon$  
by the shift $\epsilon \rightarrow \epsilon-1$. 
But as is well known that the coefficient of scalar integral $I_5^{D=6}$ 
always vanishes from the tensor integrals\cite{BDK4}. 
Thus we do not have to consider this contribution 
in the actual calculation. 
Using these results of the box integrals 
and eq.(\ref{eq:In[a]})-(\ref{eq:In[aaa]}) 
we calculate the pentagon integrals $I_5[a_i,\cdots]$ and 
other box integrals $I_4^{(k)}[a_ia_j\cdots]$.  
To perform all Feynman integrals automatically, 
we made the program of the Maple\cite{yasui}.

Now we come back to the calculation of the five gluon vertex. 
First we consider typical helicities case 
$m_5(+,+,+,+,+)$ as the simplest example. 
In this case, since the tree level amplitude vanishes  
the one-loop amplitude is infrared finite. 
We have variety of  choices of the reference momentums 
for the external gluon fields. 
We choose a reference momentum $k_i$ 
of a $i$-th gluon with a momentum $l_i$ as,
\be
\epsilon_{\mu}(l_i,k_i=l_{i+1})
={
\langle l_i|\gamma_\mu|l_{i+1}\rangle
\over
\sqrt{2}\langle l_{i+1}l_i\rangle}, 
\label{eq:5PLUS}
\ee
where $l_{i+1}$ is a ($i$+1)-th gluon momentum.  
Using the identities,
\[
l_i^{\mu_i} \varepsilon_{\mu_i}^+(l_i,k=l_{i+1})
=
l_{i+1}^{\mu_i} \varepsilon_{\mu_i}^+(l_i,k=l_{i+1})
=0,
\]
the tensor integrals can be replaced as following
,
\bea
I_5[k_1^{\alpha}k_2^{\beta}k_3^{\sigma}k_4^{\delta}\cdots]
&=&
i\int{d^Dk\over\pi^{D/2}}
{k_1^{\alpha}k_2^{\beta}k_3^{\sigma}k_4^{\delta}\cdots\over
k_1^2 k_2^2 k_3^2 k_4^2 k_5^2}
\nonumber\\
&~&
\nonumber\\
&=&\sum_{i,j,k,cdots=1}^4 
\tilde{p}_i^{\alpha}
\tilde{p}_j^{\beta}
\tilde{p}_k^{\delta}
\tilde{p}_l^{\sigma}
\cdots
I_5[a_{i+1}a_{j+1}a_{k+1}a_{l+1}\cdots]
\nonumber\\
&-&{1\over2}\sum_{ij\cdots} 
\{g\tilde{p}_i\tilde{p}_j\cdots\}^{\alpha \beta\sigma\delta \cdots}
I_5^{D=6-2\epsilon}[a_{i+1}a_{j+1}\cdots]
\nonumber\\
&+&{1\over4}\{gg\cdots\}^{\alpha \beta\sigma\delta \cdots}
I_5^{D=8-2\varepsilon}[\cdots]
\label{eq:5tensor2}
\eea
here we ignore the terms which disappear 
by contracting gluon polarization vectors eq.(\ref{eq:5PLUS}). 
$\tilde{p}_i$ are given in the table.1. 
Other combinations on $k$'s are also given by the 
permutations of $\tilde{p}$. 
\begin{center}
Table.1

\begin{tabular}{c|ccccc}\hline
i & 1 & 2 & 3 & 4 &5 \\
\hline
$\tilde{p}^\alpha_i$&
0&0&$l_3^\alpha$&$-l_5^\alpha$&0\\
$\tilde{p}^\beta_i$&
0&0&0&$l_4^\beta$&$-l_1^\beta$\\
$\tilde{p}^\sigma_i$&
$-l_2^\sigma$&0&0&0&$l_5^\sigma$\\
$\tilde{p}^\delta_i$&
$l_1^\delta$&$-l_3^\delta$&0&0&0\\
$\tilde{p}^\rho_i$&
0&$l_2^\rho$&$-l_4^\rho$&0&0\\ 
\hline
\end{tabular}
\end{center}
In addition, 
from the identities eq.(\ref{eq:<p|k|q>})$\sim$ (\ref{eq:TRACE})  
and the  expression eq.(\ref{eq:5PLUS}) of the external gluons, 
tensor structures are replaced into traces of the $\gamma$ matrices. 
For example, contraction of the momentum $l$'s and 
metric tensor $g_{\mu\nu}$ with  external gluon fields are, 
\bea
l_i^\alpha l_j^\beta l_k^\sigma l_l^\delta l_m^\rho
\times
\varepsilon_\alpha^{1+}
\varepsilon_\beta^{2+}
\varepsilon_\sigma^{3+}
\varepsilon_\delta^{4+}
\varepsilon_\rho^{5+}
&=&
{Tr(1i2j3k4l5m
P_+
)
\over(\sqrt{2})^5
\langle 21 \rangle
\langle 32 \rangle
\langle 43 \rangle
\langle 54 \rangle
\langle 15 \rangle
}
\nonumber\\
l_i^\alpha l_j^\beta l_k^\sigma g^{\delta \rho}
\times
\varepsilon_\alpha^{1+}
\varepsilon_\beta^{2+}
\varepsilon_\sigma^{3+}
\varepsilon_\delta^{4+}
\varepsilon_\rho^{5+}
&=&
{-2Tr(1i2j3k45P_+)
\over(\sqrt{2})^5
\langle 21 \rangle
\langle 32 \rangle
\langle 43 \rangle
\langle 54 \rangle
\langle 15 \rangle
}
\nonumber\\
&\cdots&
\nonumber,
\eea
where $\varepsilon_\mu^{i+}=\varepsilon_\mu^{+}(l_i,k=l_{i+1})$, 
$P_+ \equiv{1\over2}(1+\gamma_5)$, 
$Tr(ij\cdots)\equiv Tr(\slashs{l}_i\slashs{l}_j\cdots)$. 
To calculate these Dirac algebra,
we used the algebraic manipulation program FORM. 

One-loop Feynman diagrams which contribute to the $m_5$ 
are given in appendix. 
Summing up all contributions of these diagrams,  
we have the following simple final result of the color 
ordered MHV amplitude $m_{5;1}(+,+,+,+,+)$,
\[
m_{5;1}(+,+,+,+,+)= 
{i(\sqrt{2})^5N\over 96\pi^2}
{s_{12}s_{23}+s_{23}s_{34}+s_{34}s_{45}+s_{45}s_{51}+\epsilon(1,2,3,4)
\over
\langle 12 \rangle
\langle 23 \rangle
\langle 34 \rangle
\langle 45 \rangle
\langle 51 \rangle
},
\]
where $s_{ij}=2l_i\cdot l_j$ and  $\epsilon(ijkm)=i4\epsilon_{\mu\nu\rho\sigma}
l^\mu_i l^\nu_j l^\rho_k l^\sigma_m
$. 
We also applied this program to the  NHV amplitude. 
For the NHV amplitude $m_{5;1}(-,+,+,+,+)$, 
we chose the reference momenta $k_i$ 
as $(k_1,k_2,k_3,k_4,k_5)=(l_2,l_3,l_4,l_5,l_2)$. 
The NHV amplitude $m_{5;1}(-,+,+,+,+)$ is given by,
\[
m_{5;1}(-,+,+,+,+)= {i(\sqrt{2})^5N\over 96\pi^2}
{A + B\times\epsilon(1,2,3,4) \over 
[12][51]
\langle 23\rangle
\langle 34\rangle
\langle 45\rangle
\langle 25\rangle^2
},
\]
where
\bea
A&=&
\left (s_{{34}}+s_{{23}}\right ){s_{{12}}\over s_{15}}\left \{{\frac 
{{s_{{45}}}^{2}
s_{{34}}}{s_{{23}}}}+{\frac {{s_{{12}}}^{2}s_{{23}}}{s_{{34}}}}+2\,s_{{
12}}s_{{45}}+s_{{23}}s_{{34}}-2\,s_{{34}}s_{{45}}-2\,s_{{23}}s_{{12}}
\right \}
\nonumber\\
&+&
\left (s_{{34}}+s_{{45}}\right ){s_{{15}}\over s_{12}}\left \{{\frac 
{{s_{{23}}}^{2}S
_{{34}}}{s_{{45}}}}+{\frac {{s_{{15}}}^{2}s_{{45}}}{s_{{34}}}}+2\,s_{{
15}}s_{{23}}+s_{{34}}s_{{45}}-2\,s_{{23}}s_{{34}}-2\,s_{{15}}s_{{45}}
\right \}
\nonumber\\
&-&
{s_{{12}}s_{13}\over s_{23}}\left (s_{{15}}s_{{12}}+2\,s_
{{34}}s_{{45}}-s_{{15}}s_{{45}}\right )
-
{s_{{15}}s_{14}\over s_{45}}\left (2\,s_{{23}}s_{{34}}-s_
{{23}}s_{{12}}+s_{{15}}s_{{12}}\right )
\nonumber\\
&+&{1\over s_{34}}\{
(s_{15}+s_{12})^2(s_{15}s_{12}+s_{23}s_{45}-2s_{12}s_{23}
-2s_{15}s_{45})
\nonumber\\
&~&
+(s_{15}s_{45}+s_{15}s_{23}+s_{12}s_{45}+s_{12}s_{23})(s_{15}s_{45}+s_{12}s_{23})
+s_{15}^2s_{45}^2+s_{12}^2s_{23}^2
\}
\nonumber\\
&
-&
6s_{15}s_{34}s_{45}+2s_{15}s_{23}s_{45}+s_{15}s_{34}s_{23}+2s_{12}s_{23}s_{45}
+s_{15}s_{12}s_{34}+2s_{15}^2s_{23}
\nonumber\\
&+&2s_{12}^2s_{45}+s_{34}^2s_{45}+s_{23}s_{34}^2-2s_{12}s_{23}^2+7s_{23}s_{12}^2
-2s_{15}s_{45}^2+7s_{15}^2s_{45}
\nonumber\\
&-&6s_{23}s_{12}s_{34}
+8s_{15}s_{23}s_{12}+8s_{15}s_{12}s_{45}+s_{12}s_{34}s_{45},
\nonumber
\eea
and
\bea
B&=&
{s_{{12}}\over s_{15}}\left (s_{{34}}+s_{{23}}\right )
\left \{-{\frac {s_{{45}}}{s_{{
23}}}}-{\frac {s_{{12}}}{s_{{34}}}}+1\right \}
+
{s_{{15}}\over s_{12}}\left (s_{{34}}+s_{{45}}\right )
\left \{-{\frac {s_{{23}}}{s_{{
45}}}}-{\frac {s_{{15}}}{s_{{34}}}}+1\right \}
\nonumber\\
&+&
{1\over s_{34}}\left\{
-2\,s_{{15}}s_{{45}}-s_{{15}}s_{{23}}-s_{{12}}s_{{45}}-2\,s_{{12}}s_{{
23}}+2\,s_{{15}}s_{{12}}+{s_{{12}}}^{2}+{s_{{15}}}^{2}\right\}
\nonumber\\
&+&
{s_{{12}}s_{13}\over s_{23}}
+
{s_{{15}}s_{14}\over s_{45}}
-4\,s_{{12}}-4\,s_{{15}}+s_{{34}}.
\nonumber
\eea
These results are consistent with Bern, Kosower and Dixson's results.
(The difference between this result and BDK's result 
in the factor $(\sqrt{2})^5$ come from the 
different normalization of $T^a$.)
We reaffirmed the BDK's results for the 
$m_{5;1}(-,-,+,+,+)$ case and  $m_{5;1}(-,+,-,+,+)$ case, too.

\section{Conclusion}

A combination of the background field gauge, color decomposition 
and spinor helicity basis is presented as a very powerful method for the 
loop calculations in perturbative QCD.  The method takes advantage of the
three facts:  1) The simple structure of the color ordered Feynman rule in
the  background field gauge suppresses the number of the  terms in the
intermediate stages  of the loop calculations.  
2)Using the color decomposition technique, only a
color-ordered  subset of all possible Feynman diagrams is required. 
3) Appropriate choices of the reference momenta
in the helicity basis reduces the usually complicated tensor structure
to a much simplified Dirac algebra.   This also contributes in suppressing
the number of the intermediate terms.  These advantages simplify the loop
calculation and makes multi-jet amplitude calculations at
one-loop order feasible.  This simplicity has been demonstrated in the
example of five-gluon amplitude at the one loop level \cite{BDK3}. 

Since the method is formulated solely within the conventional gauge field
theory, it is applicable to a wider range of calculations than one-loop
perturbative QCD.  For example, the extension of 
the background field gauge 
to theories with spontaneously broken symmetry 
like the electroweak theory is straightforward \cite{DENNER}.  
In this case, the Feynman rule still possesses the simple structure. 
We expect the method simplifies two-loop calculations, too.

\vspace{0.5cm}
The author would like to thank Prof.L.Trueman and Prof.S.Ohta and 
Dr. Kilgore  
for the useful discussions and  for reading the manuscript, 
and wishes to thank  
Prof. T.D.Lee, 
Prof.Z.Bern, Prof.J.Kodaira, Prof.T.Uematsu and Dr.T.Tanaka 
for the useful comments.
The author also thanks RIKEN, Brookhaven National Laboratory and the 
U.S. Department of Energy  for providing the facilities essential for
the completion of this work.

\appendix

\section{Feynman diagrams for the one-loop five gluon vertex}
Here we give the Feynman diagrams which contribute to the one-loop level 
five gluon amplitudes. We only show the gluonic loop diagrams. 
In the actual calculation, we need the ghost loop diagrams, too.
We do not consider the renormalization of the wave functions.
In the dimensional regularization, the diagrams 9) and 13) 
give no contribution. 

\vspace{0.5cm}
\raisebox{2.7cm}{1)}
\resizebox{3cm}{3cm}{\includegraphics{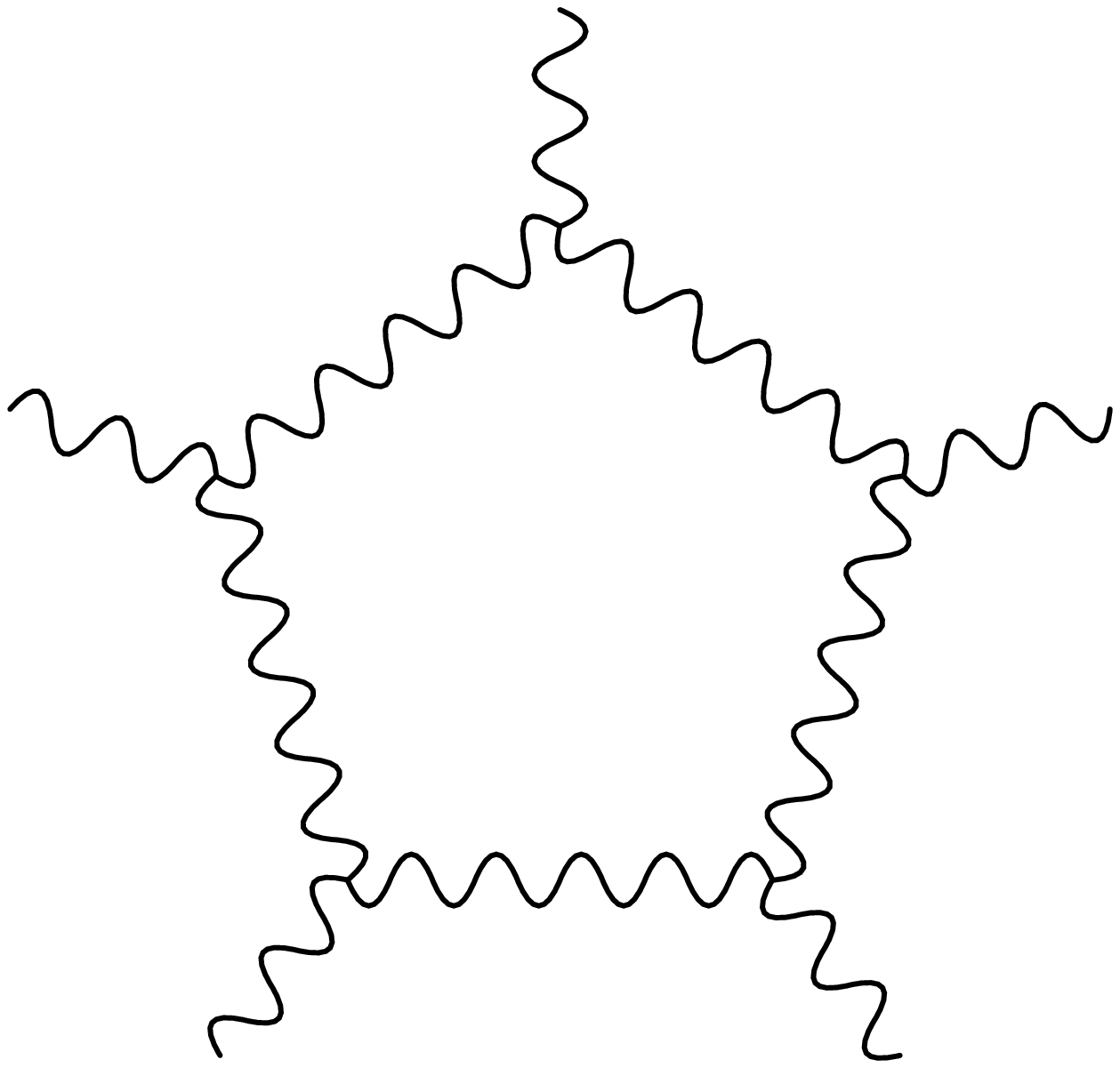} }
\raisebox{2.7cm}{2)}
\resizebox{3cm}{3cm}{\includegraphics{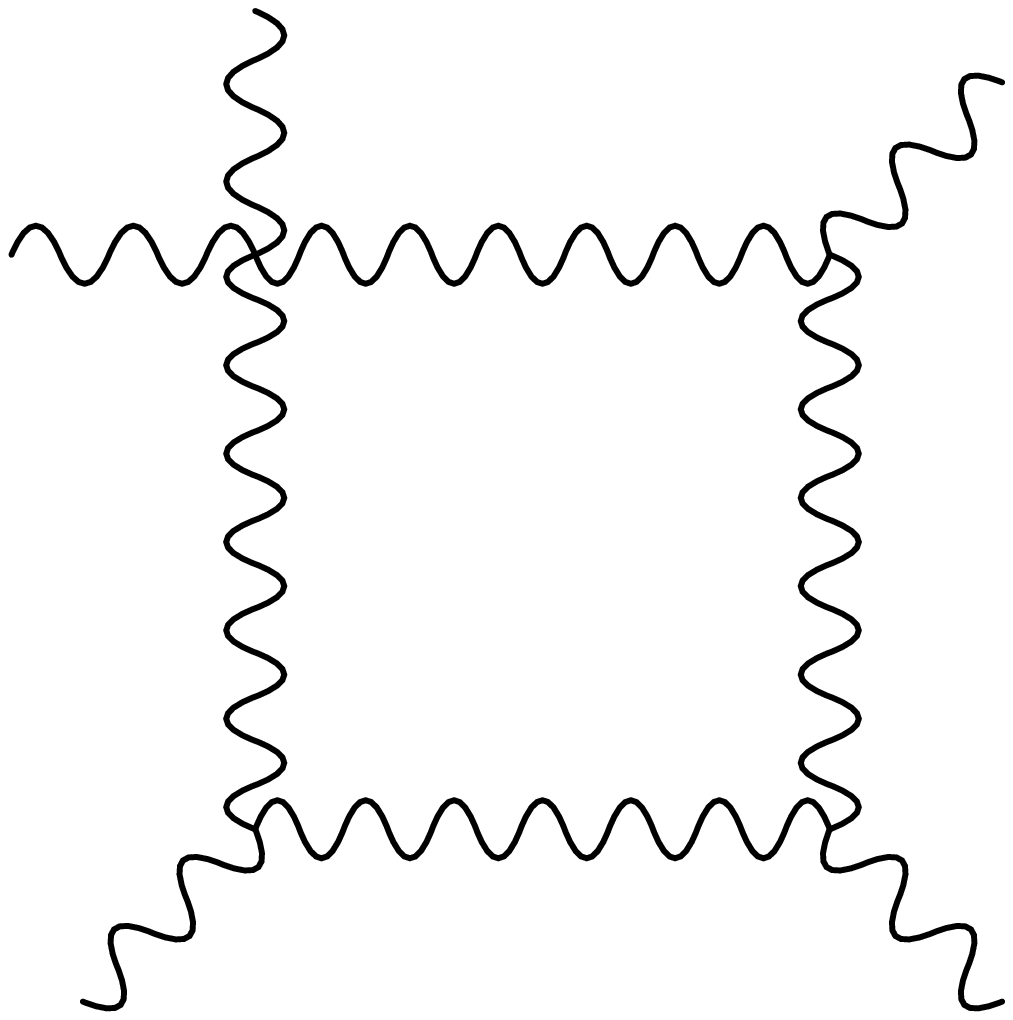} }
\raisebox{2.7cm}{3)}
\resizebox{3cm}{3cm}{\includegraphics{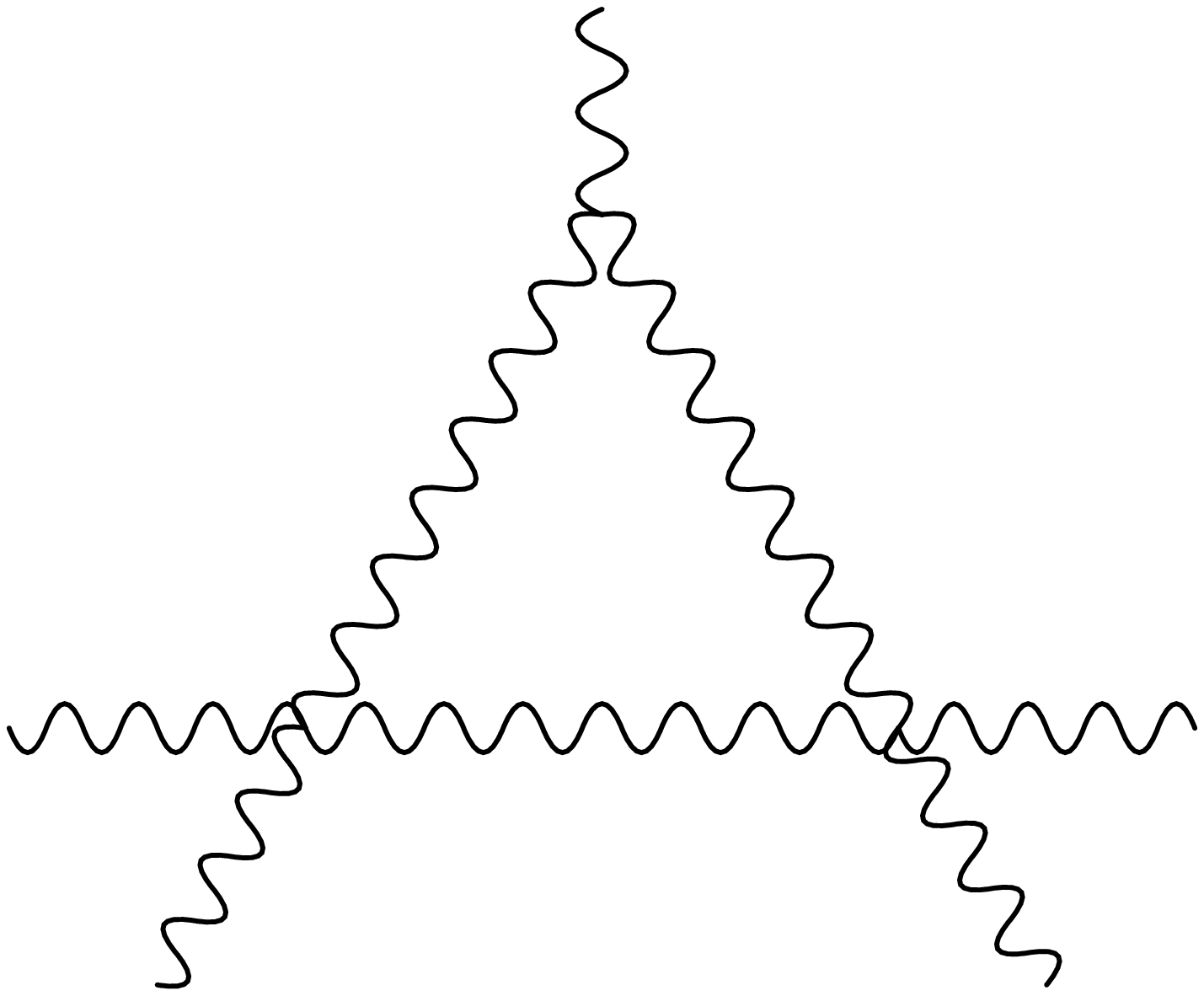} }

\vspace{0.5cm}
\raisebox{2.7cm}{4)}
\resizebox{3cm}{3cm}{\includegraphics{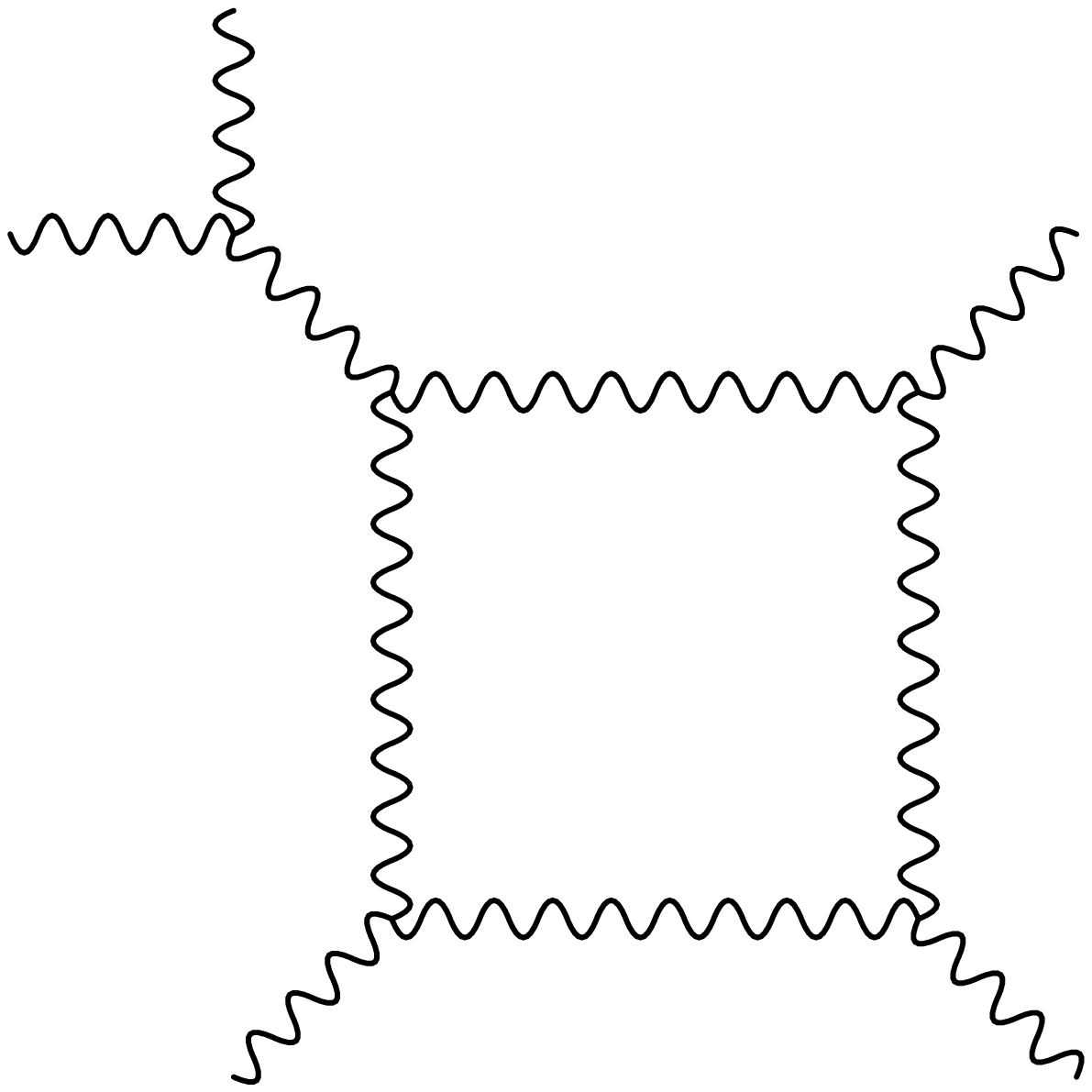} }
\raisebox{2.7cm}{5)}
\resizebox{3cm}{3cm}{\includegraphics{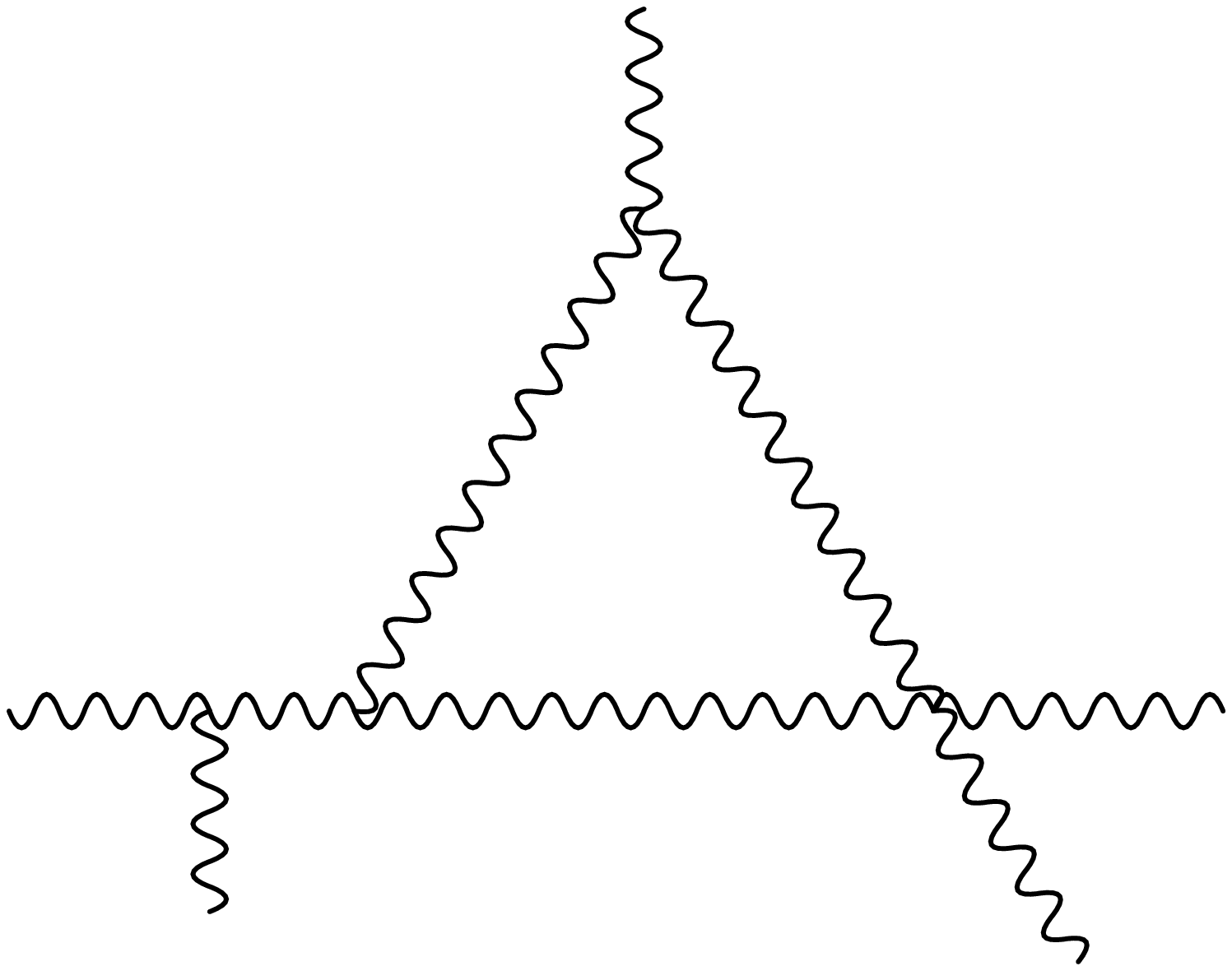} }
\raisebox{2.7cm}{6)}
\resizebox{3.5cm}{3cm}{\includegraphics{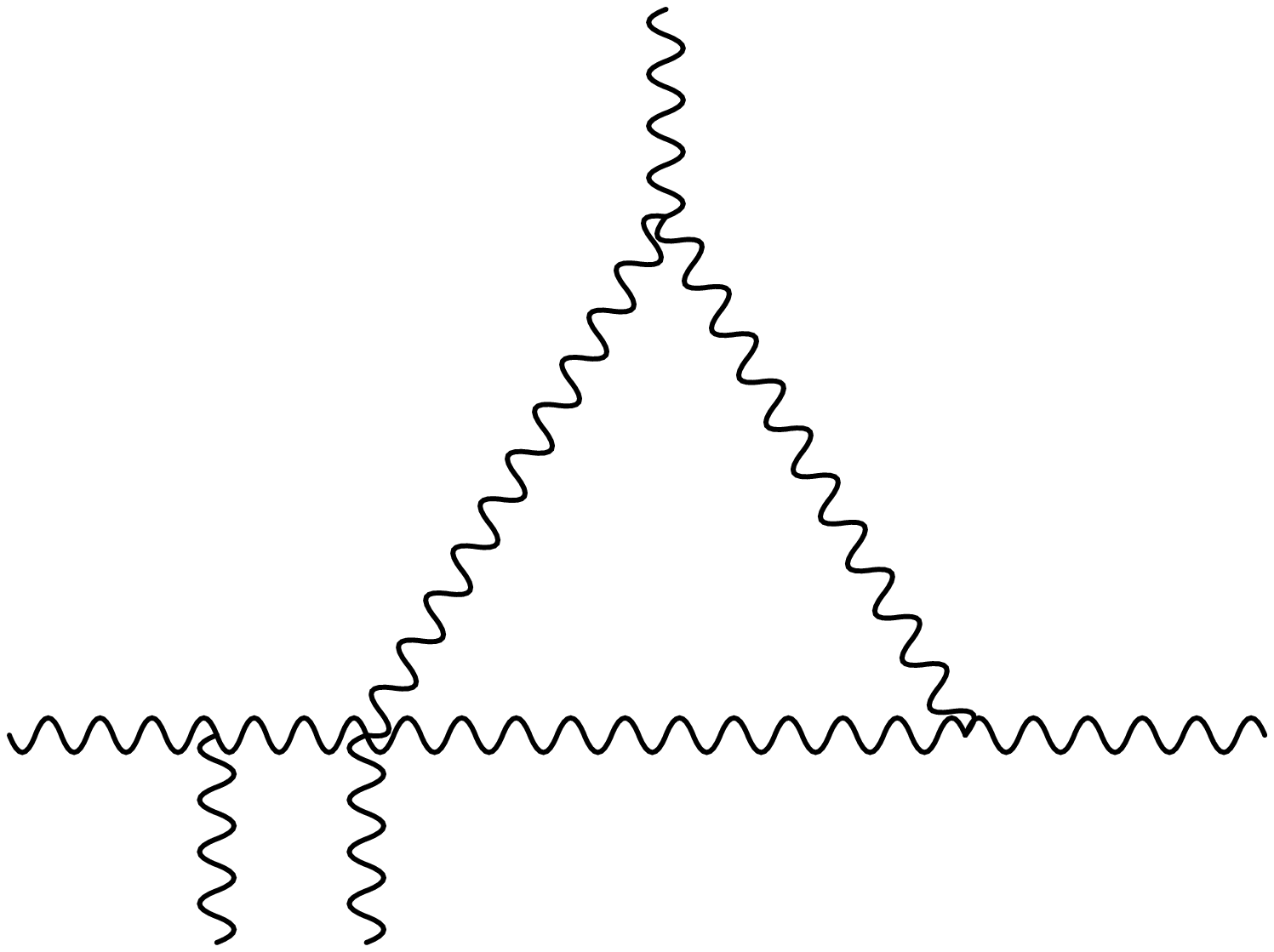} }
\raisebox{2.7cm}{7)}
\resizebox{3.5cm}{2.5cm}{\includegraphics{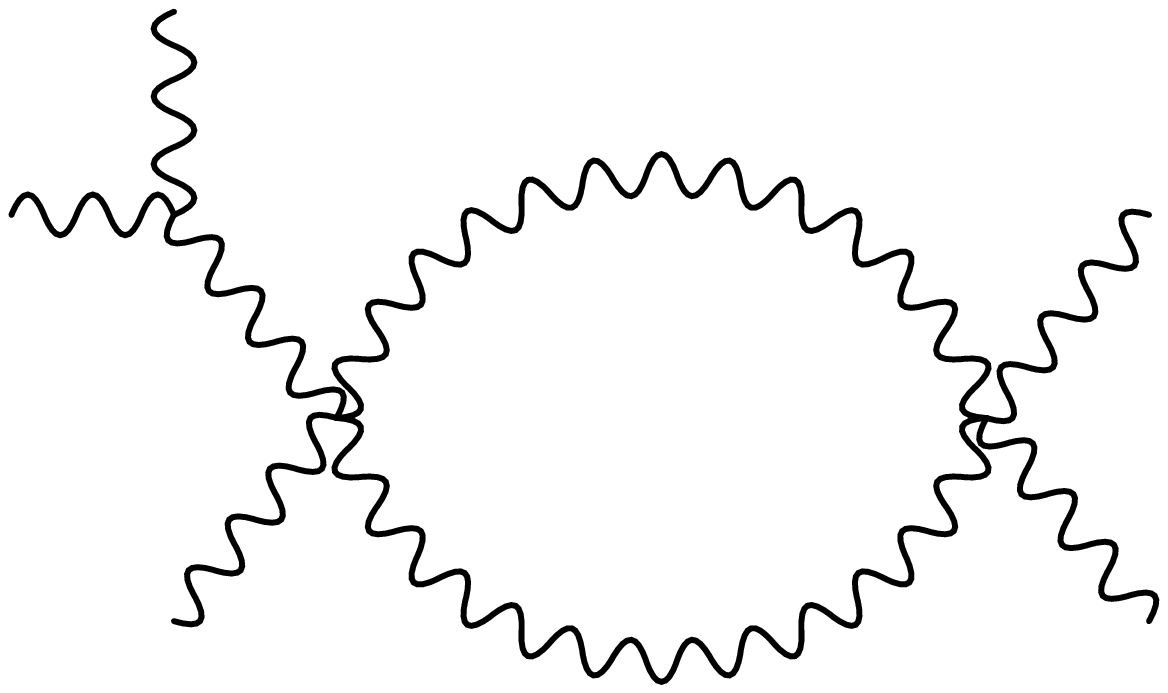} }

\vspace{0.5cm}
\raisebox{2.7cm}{8)}
\resizebox{3cm}{3cm}{\includegraphics{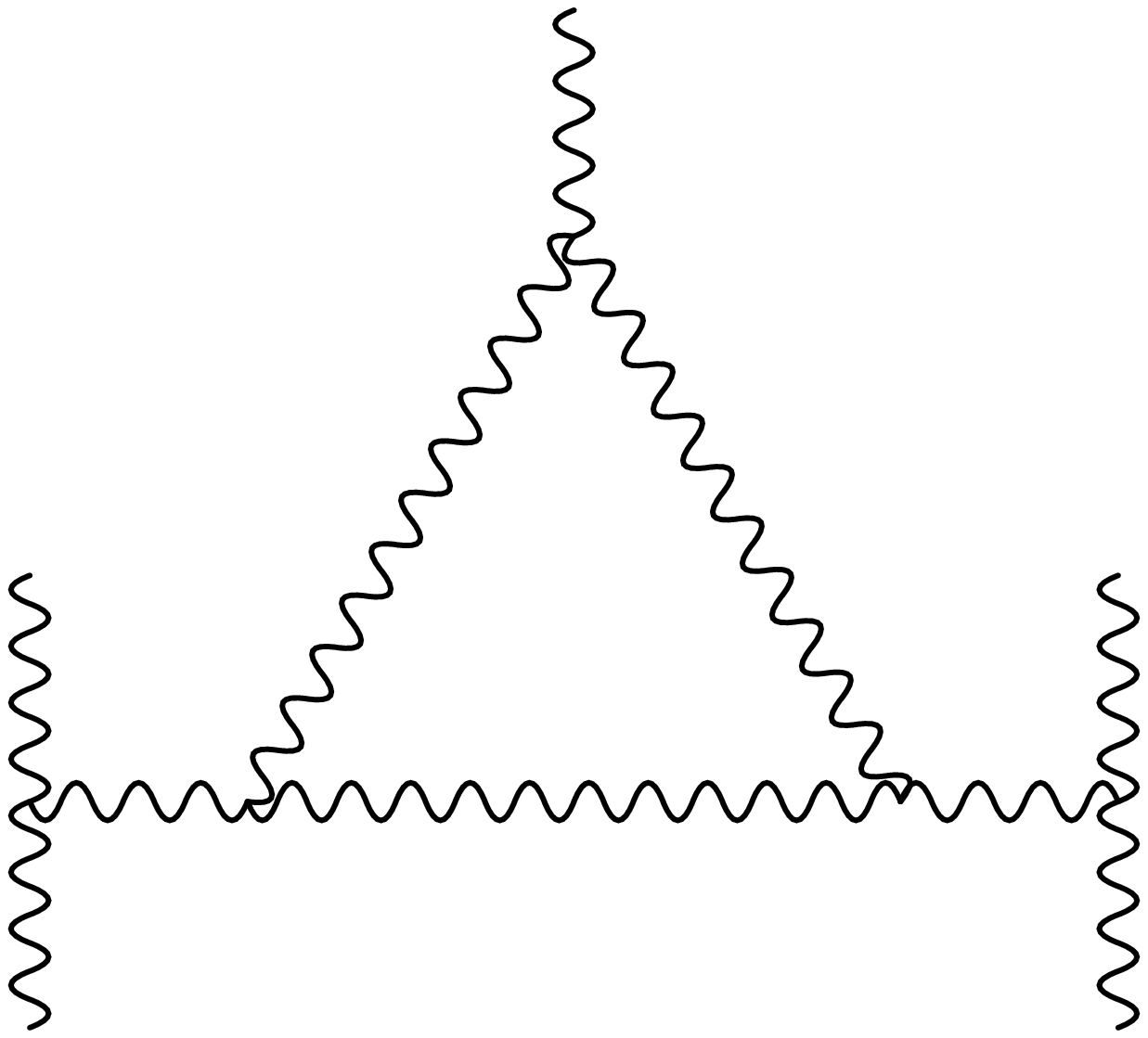} }
\raisebox{2.7cm}{9)}
\resizebox{3cm}{3cm}{\includegraphics{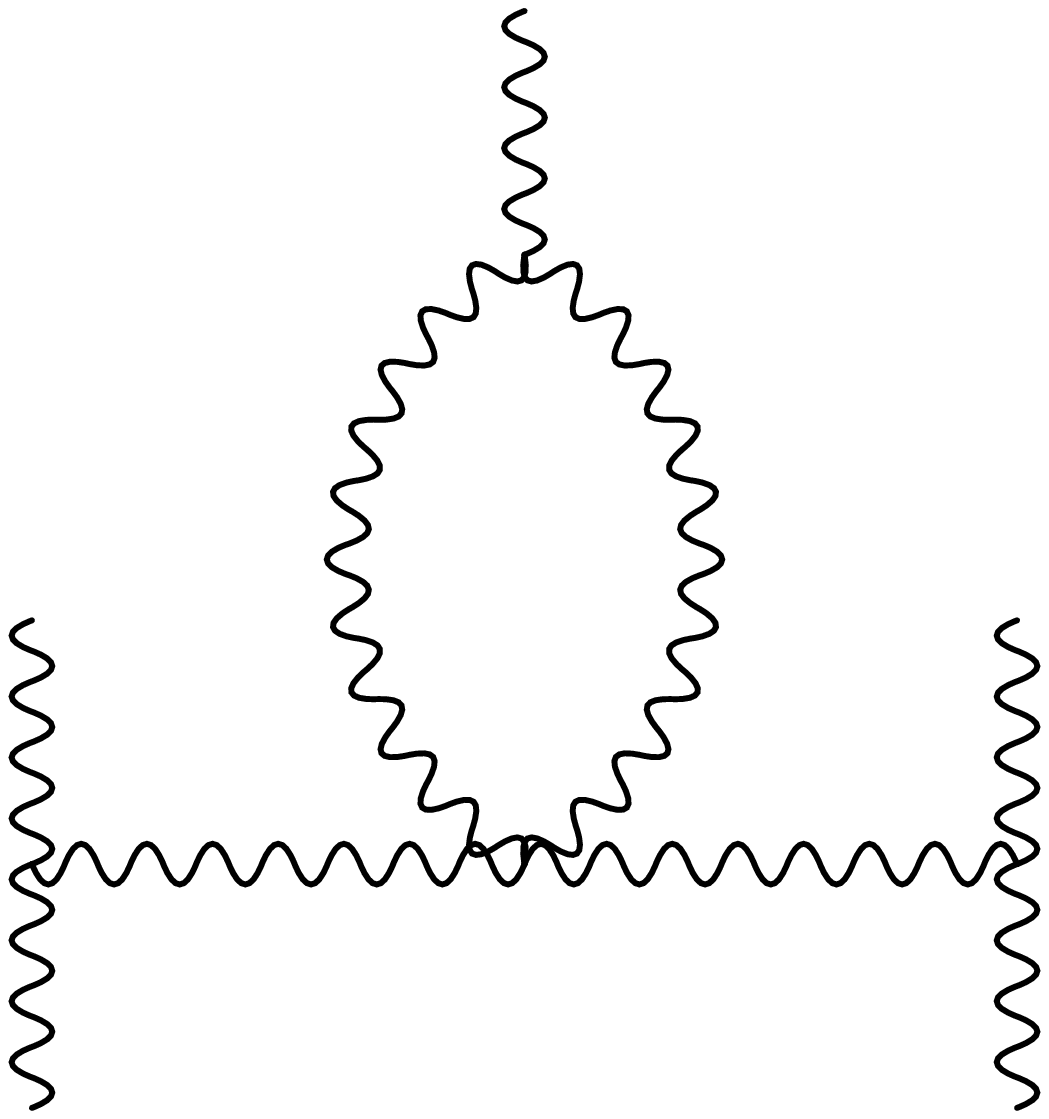} }
\raisebox{2.7cm}{10)}
\resizebox{3cm}{2.5cm}{\includegraphics{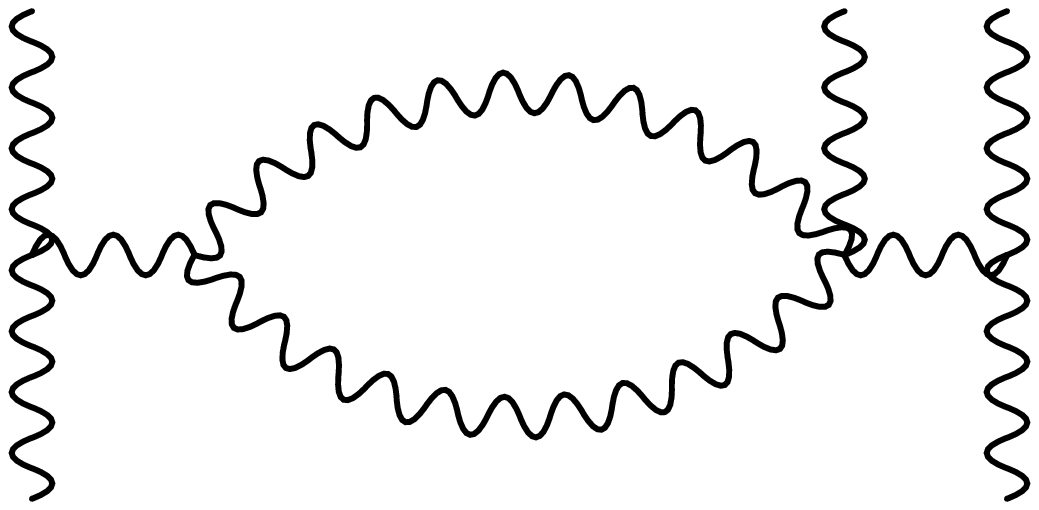}}

\vspace{0.5cm}
\raisebox{2.7cm}{11)}
\resizebox{3.5cm}{2cm}{\includegraphics{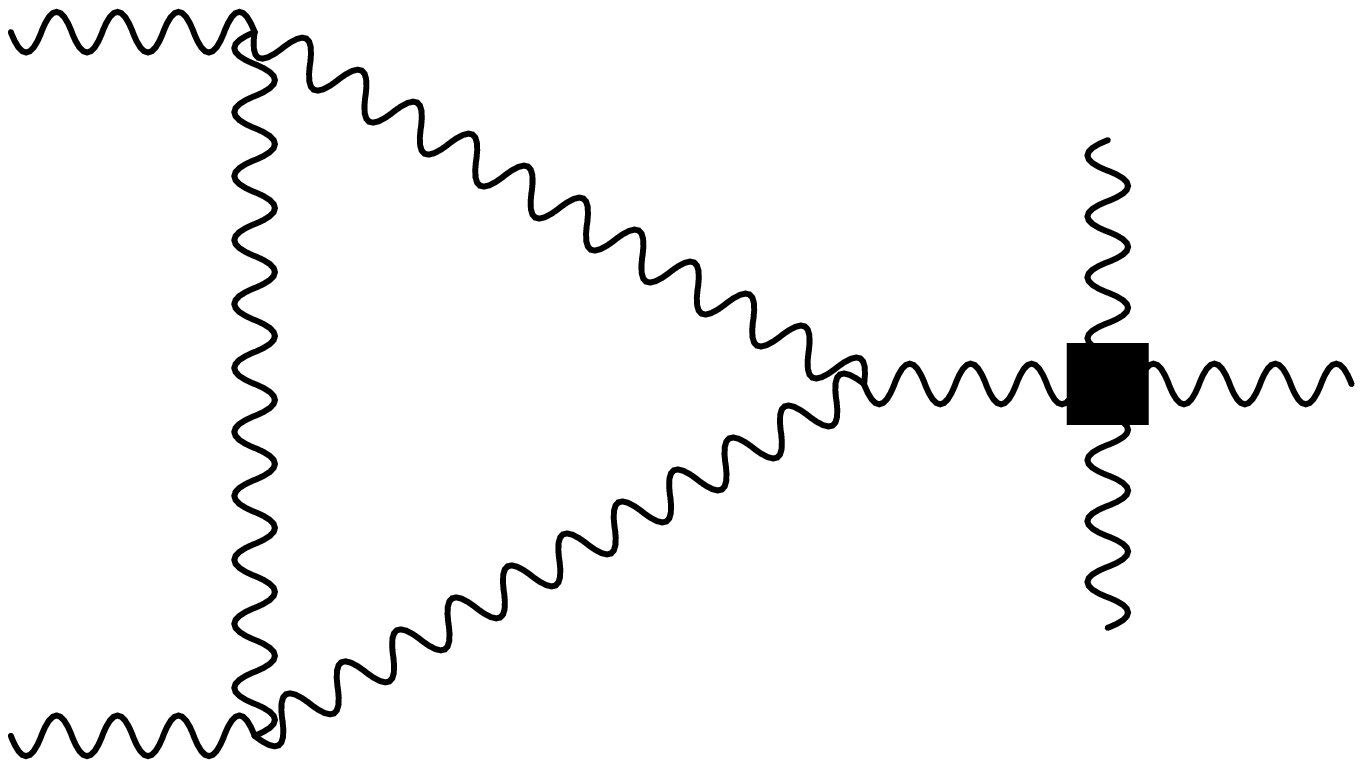} }
\raisebox{2.7cm}{12)}
\resizebox{3.5cm}{2cm}{\includegraphics{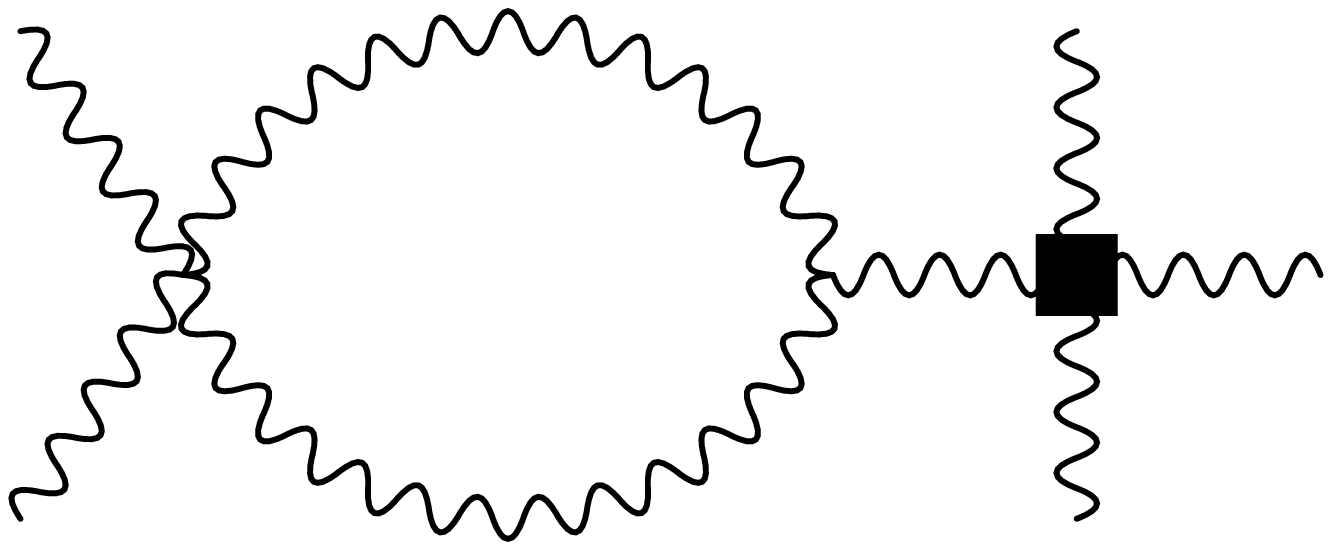} }
\raisebox{2.7cm}{13)}
\resizebox{3.5cm}{2cm}{\includegraphics{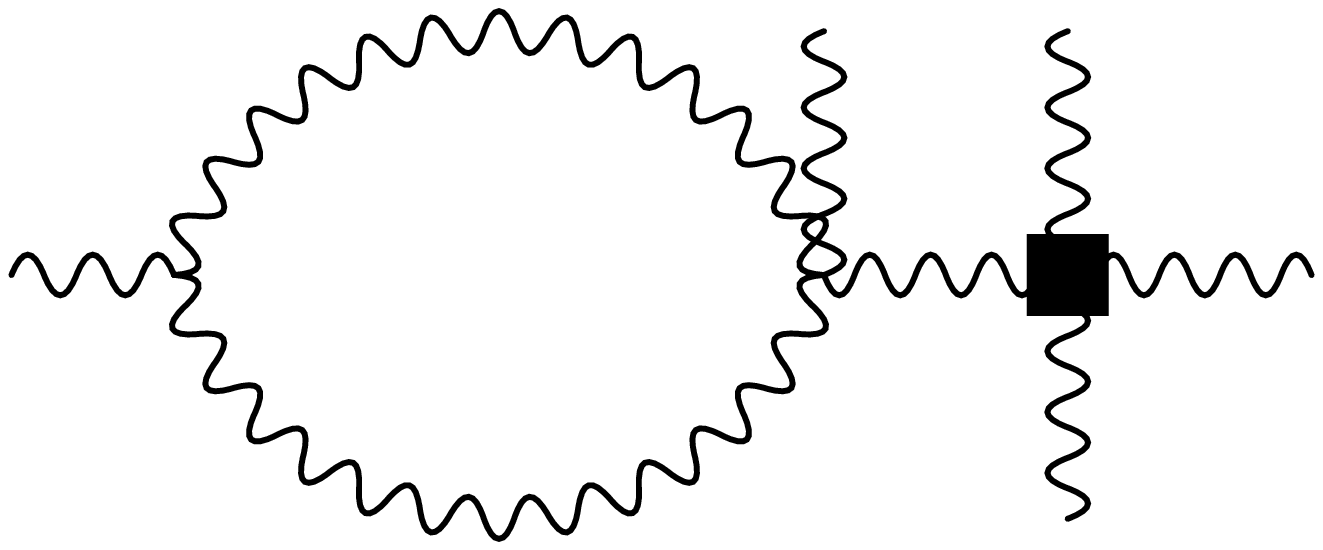} }
\raisebox{2.7cm}{14)}
\resizebox{3.5cm}{2cm}{\includegraphics{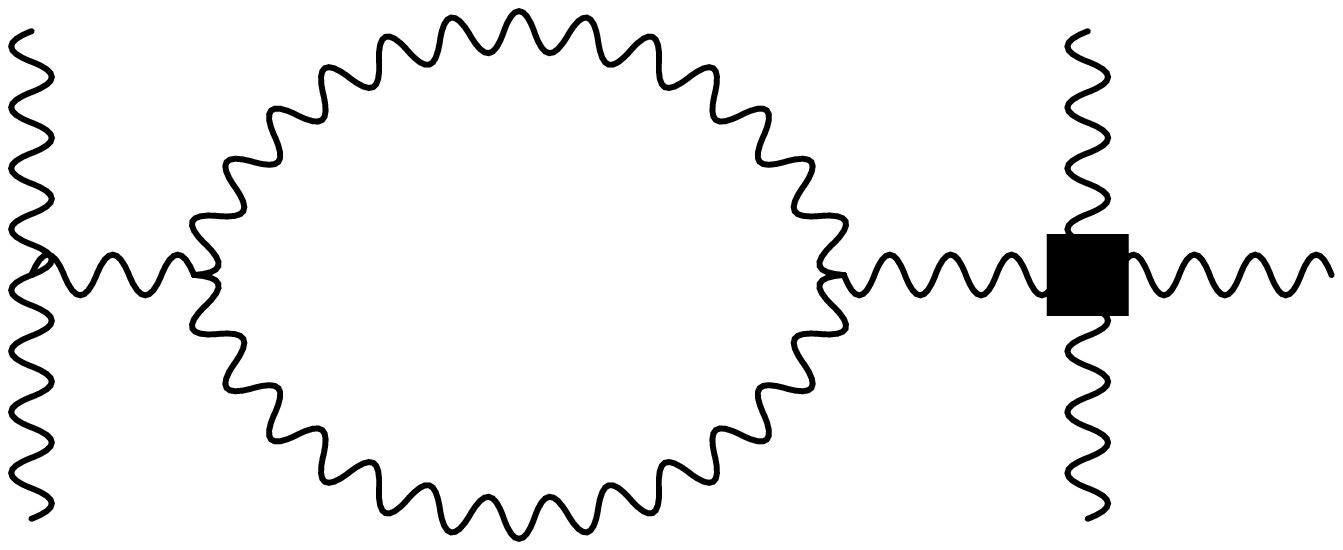} }

\vspace{1cm}
Here, 
\vspace{0.3cm}

\resizebox{2cm}{2cm}{\includegraphics{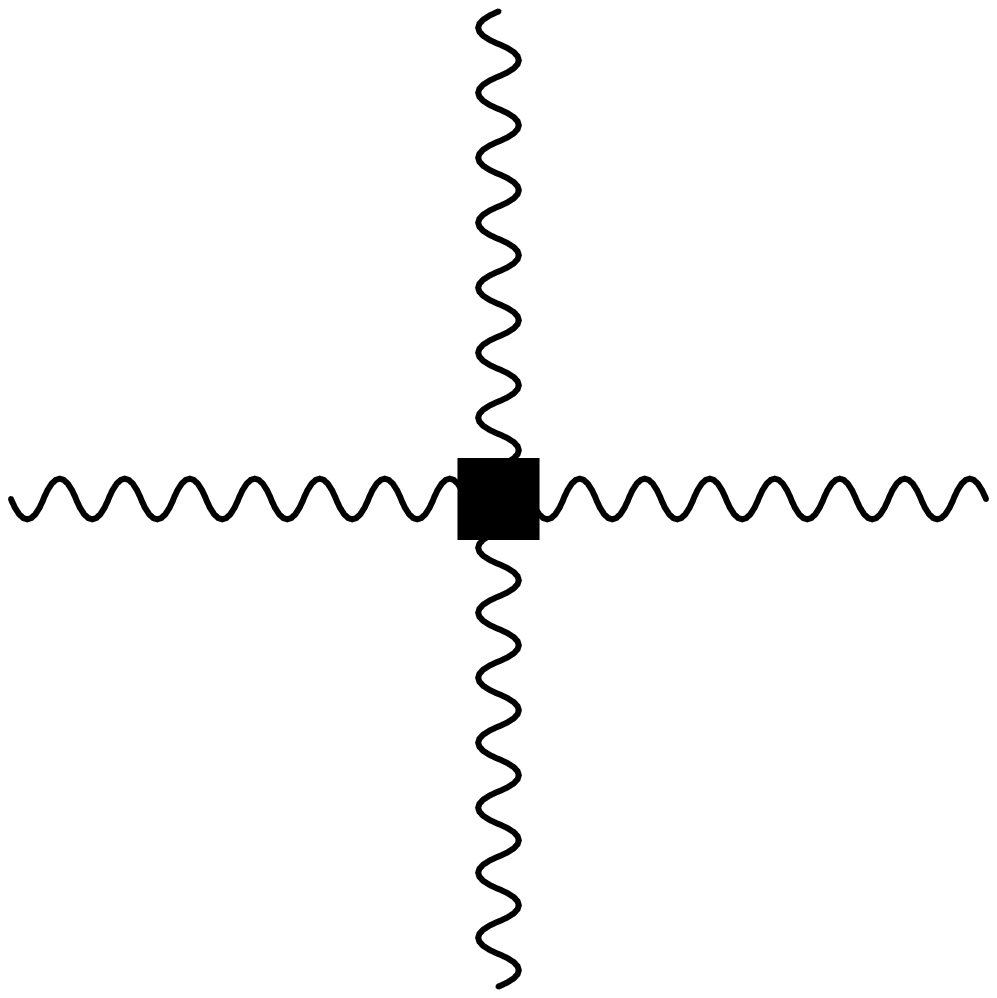} }
\raisebox{1cm}{=}
\resizebox{2.5cm}{2cm}{\includegraphics{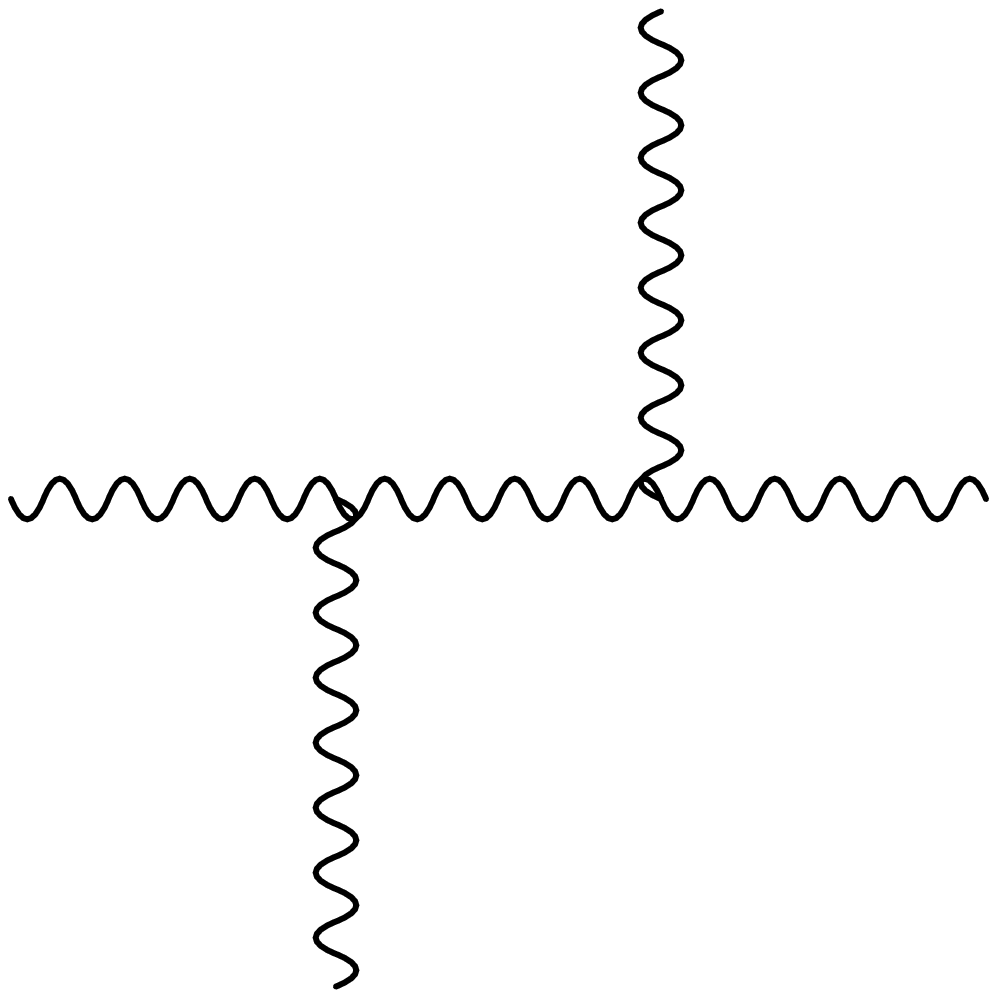} }
\raisebox{1cm}{+}
\resizebox{2.5cm}{2cm}{\includegraphics{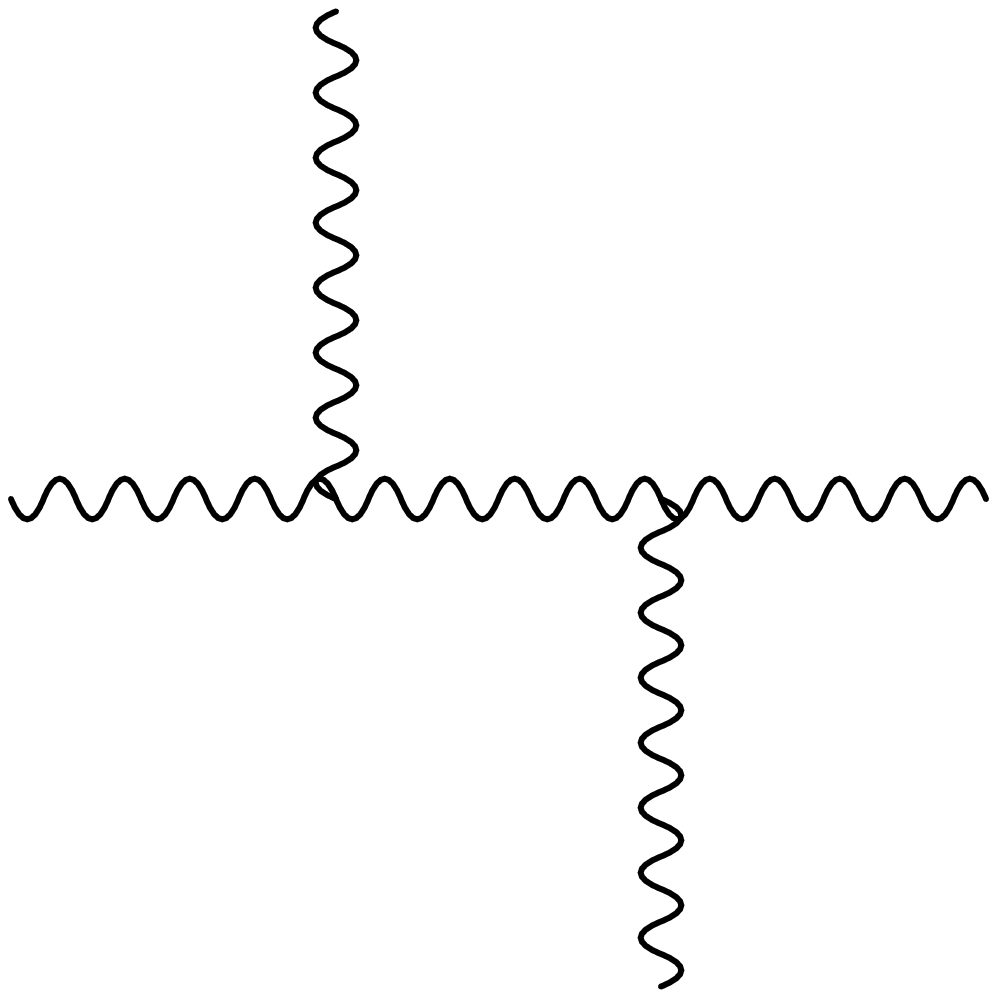} }
\raisebox{1cm}{+}
\resizebox{2.5cm}{2cm}{\includegraphics{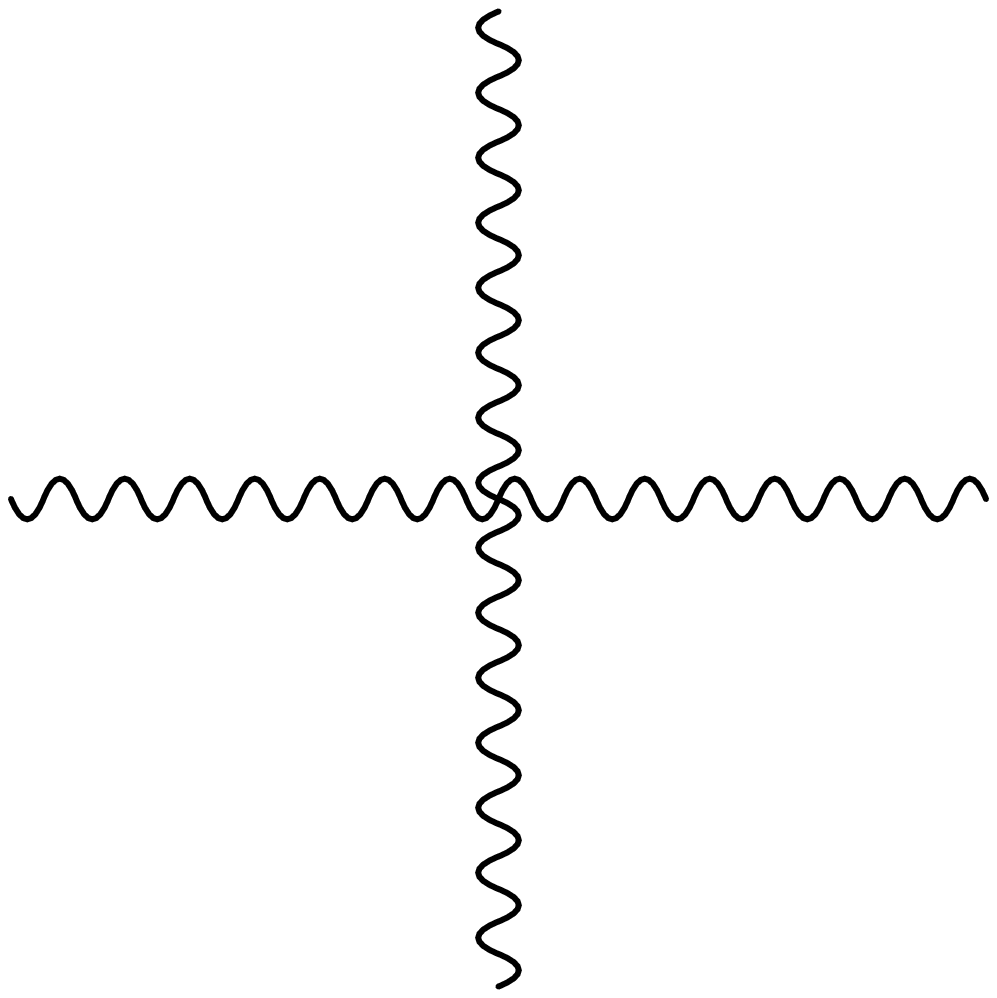} }


\begin{thebibliography}{99}

\bibitem{PARKE1}
M. Mangano and S. Parke, {\sl Phys. Rep.}{\bf 200}301 (1991), 
and reference therein.

\bibitem{color}
J.E. Paton and H.-M.Chan, {\sl Nucl. Phys.}
{\bf B10} 516 (1969);\\
P.Cvitanovic, P.G.Lauwers, and P.N.Scharbach {\sl Nucl. Phys.}
{\bf B186} 165 (1981);\\
F. A. Berends and W.T. Giele, {\sl Nucl. Phys.}
{\bf B294} 700 (1987);\\
D.A.Kosower, B.-H.Lee and V.P.Nair, {\sl Phys. Lett.}
{\bf B201} 85 (1988);\\
M. Mangano, S.Parke and Z. Xu, {\sl Nucl. Phys.}
{\bf B298} 653 (1988);\\
M. Mangano, {\sl Nucl. Phys.}
{\bf B309} 461 (1988);\\
D.Zeppenfeld,  {\sl Int.J.Mod.Phys.}
{\bf A 3} 2175 (1988);\\
Z.Bern and D.A.Kosower,{\sl Nucl. Phys.}
{\bf B362} 389 (1991).


\bibitem{SHB}
F.A.Bernds, R. Kleiss, P.De Causmsecker, 
R. Gastmans, and T.T.Wu, {\sl Phys. Lett}
{\bf 103B} 124 (1981);\\
P.De Causmsecker,R. Gastmans, W.Troost, and T.T.Wu,
 {\sl Nucl. Phys.}{\bf B206} 53 (1982);\\
J. Gunion and Z. Kunszt,  {\sl Phys. Lett.}{\bf B161} 333 (1985);\\
R. Kleiss and W.J. Stirling, {\sl Nucl. Phys.}{\bf B262} 235 (1985);\\
Z.Xu, Da-Hua Zhang and L. Chang, 
 {\sl Nucl. Phys.}{\bf B291} 392(1987).

\bibitem{PARKE2}
M.T.Grisaru, H.N.Pendleton and P.van Nieuwenhuizen,
{\sl Phys. Rev.} {\bf D15} 996 (1977);\\
M.T.Grisaru and H.N.Pendleton
{\sl Nucl. Phys.} {\bf B124} 81 (1977);\\
S. Parke and T. Tayler, {\sl Phys. Lett.}{\bf B157} 81 (1985).


\bibitem{BDK1}
Z.Bern and D.A. Kosower, {\sl Phys. Rev. Lett.} {\bf 66} 1669(1991).

\bibitem{BDK2}
Z.Bern and D.A. Kosower, {\sl Nucl. Phys.}{\bf B379} 451(1992).

\bibitem{BDK3}
Z.Bern, L. Dixon  and D.A. Kosower, {\sl Phys. Rev. Lett.}{\bf 70} 
2677 (1993).

\bibitem{ELLIS}
R.K.Ellis and J.C.Sexton, {\sl Nucl. Phys.}{\bf B269} 445 (1986).


\bibitem{KILG}
W.B. Kilgore, and W.T. Giele 
{\sl Phys. Rev.} {\bf D55} 7183 (1997).

\bibitem{BD}
Z.Bern, D.C.Dunbar {\sl Nucl. Phys.}{\bf B379} 562 (1992).

\bibitem{BK}
Z.Bern and D.A.Kosower, {\sl Nucl. Phys.}{\bf B362} 389 (1991).


\bibitem{DRM}
        J.Scherk, {\sl Nucl. Phys.} {\bf B31} 222 (1971);\\ 
        A. Neveu and J.Scherk, 
           {\sl Nucl. Phys.} {\bf B36} 155 (1971);\\
        J.L. Gervais andA. Neveu, 
           {\sl Nucl. Phys.} {\bf B42} 381 (1972).


\bibitem{DEWITT}
      B. S. DeWitt, {\sl Phys. Rev.} {\bf 162} 1195 (1967);\\
      Dynamical Theory of Groups and Fields 
      (Gordon and Breach, New York,1963);\\
      and also in Quantum Gravity 2, edited by C.J.Isham, R.Penrose 
      and D. S. Sciama 
      (Oxford University Press, New York, 1981).

\bibitem{HART}
     C.F. Hart,{\sl Phys. Rev. }{\bf D28} 1993 (1983).
\bibitem{ABB}
     L.F. Abbott, {\sl Nucl. Phys.} {\bf B185} 189(1981);\\
     {\sl Acta Phys. Pol.} {\bf B13} (1982) 33.  

\bibitem{ABB2}
     L.F. Abbott, {\sl Nucl. Phys.} {\bf B229} (1983) 372.


\bibitem{KOS}
D. Kosower, {\sl Phys. Lett.}{\bf B254} 439 (1991).

\bibitem{BDK4}
Z.Bern, L. Dixon  and D.A. Kosower, {\sl Phys. Lett.}{\bf B302} 299 (1993);\\
Z.Bern, L. Dixon  and D.A. Kosower, {\sl Nucl.Phys.} {\bf B412} 751(1994).


\bibitem{BDK5}
Z. Bern, L. Dixon, D. Dunbar and D.A.Kosower,
{\sl Nucl.Phys.} {\bf B425} 217 (1994).

\bibitem{VAN}
D.B. Melrose, {\sl Il Nuovo Cimento} {\bf 40A} 181 (1965);\\ 
G. J. van Neerven and J.A.M. Vermaseren, {\sl Phys.Lett.}
{\bf B137} 241 (1984).

\bibitem{thooft}
G. 'tHooft and M.Veltman, {\sl Nucl.Phys.} {\bf B153} 365 (1979).

\bibitem{yasui} Y. Yasui, in preparation

\bibitem{DENNER}
A. Denner, S. Dittmaier and G. Weiglein {\sl Acta. Phys. Polon.}
{\bf B27} 3645 (1996).


\end{thebibliography}
\end{document}